\documentclass{ieeeaccess}
\usepackage{cite}
\usepackage{textcomp}
\usepackage{graphicx}
\usepackage{amsmath,amssymb,amsfonts}
\usepackage{multirow}
\usepackage{psfrag}
\usepackage{subfigure}
\usepackage{color}
\usepackage{soul}
\usepackage{xcolor}
\usepackage{lipsum}
\usepackage[ruled,vlined,linesnumbered]{algorithm2e}
\usepackage{mathrsfs}
\usepackage{pbox}

\def\BibTeX{{\rm B\kern-.05em{\sc i\kern-.025em b}\kern-.08em
    T\kern-.1667em\lower.7ex\hbox{E}\kern-.125emX}}

\graphicspath{{Figures/}}

\begin{document}
\history{Date of publication xxxx 00, 0000, date of current version xxxx 00, 0000.}
\doi{10.1109/ACCESS.2017.DOI}

\title{Intelligent Traffic Monitoring Systems for Vehicle Classification: A Survey}
\author{\uppercase{Myounggyu Won}\authorrefmark{1}, \IEEEmembership{Member, IEEE}}
\address[1]{Department of Computer Science, University of Memphis, Memphis, TN 38152, USA}
\tfootnote{This research was supported in part by the Competitive Research Grant Program (CRGP) of South Dakota Board of Regents (SDBoR), and in part by Global Research Laboratory Program (2013K1A1A2A02078326) through NRF, and DGIST Research and Development Program (CPS Global Center) funded by the Ministry of Science, ICT \& Future Planning of South Korea.}

\markboth
{Author \headeretal: Preparation of Papers for IEEE TRANSACTIONS and JOURNALS}
{Author \headeretal: Preparation of Papers for IEEE TRANSACTIONS and JOURNALS}

\corresp{Corresponding authors: Myounggyu Won (email: mwon@memphis.edu).}

\begin{abstract}
A traffic monitoring system is an integral part of Intelligent Transportation Systems (ITS). It is one of the critical transportation infrastructures that transportation agencies invest a huge amount of money to collect and analyze the traffic data to better utilize the roadway systems, improve the safety of transportation, and establish future transportation plans. With recent advances in MEMS, machine learning, and wireless communication technologies, numerous innovative traffic monitoring systems have been developed. In this article, we present a review of state-of-the-art traffic monitoring systems focusing on the major functionality--vehicle classification. We organize various vehicle classification systems, examine research issues and technical challenges, and discuss hardware/software design, deployment experience, and system performance of vehicle classification systems. Finally, we discuss a number of critical open problems and future research directions in an aim to provide valuable resources to academia, industry, and government agencies for selecting appropriate technologies for their traffic monitoring applications.
\end{abstract}


\titlepgskip=-15pt

\maketitle

\section{Introduction}
\label{sec:introduction}

As the number of vehicles has increased significantly, the capacity of existing transportation networks is almost at its maximum, causing severe traffic congestion in many countries~\cite{won2016toward}. Constructing additional highway infrastructure, however, is not a feasible option because of the high cost and limited space. For example, constructing a high occupancy vehicle (HOV) lane in the city of Los Angeles costs up to \$750,000 per lane and per mile~\cite{2016TrafficGuide}. In particular, the expenses increase prohibitively to provide safety to construction workers and build extra facilities to maintain traffic flow during construction.

A traffic monitoring system is an effective alternative to mitigate traffic congestion. It is an integral component of Intelligent Transportation Systems (ITS) that is used to collect traffic data such as the number of vehicles, types of vehicles, and vehicle speed. Based on the collected data, it performs traffic analysis to better utilize the roadway systems, predict future transportation needs, and improve the safety of transportation~\cite{won2018deepwitraffic}. Transportation agencies in many countries spend huge amounts of money to develop, deploy, and maintain traffic monitoring systems~\cite{lee2015using}.

One of the main functionalities of a traffic monitoring system is vehicle classification. Especially, due to significant technical challenges, various research issues have been investigated regarding vehicle classification leading to development of numerous vehicle classification systems. Classifying vehicles into different types accurately is of crucial importance for effective traffic operation and transportation planning. For example, the information about the number of large trucks on a highway section is used to estimate the capacity of the highway section and plan for pavement maintenance work. Identifying the vehicle types especially the number of multi-unit vehicles is of a great interest to the safety community. Even the geometric roadway design is dictated by the vehicle types that frequently utilize the roadway.

Numerous vehicle classification systems have been developed. Especially, recent advances in sensing, machine learning, and wireless communication technologies gave rise to numerous innovative vehicle classification systems. Although these new classification systems enable vehicle classification with higher accuracy, they have significantly different characteristics and requirements such as types of sensors used, hardware settings, configuration process, parameter settings, operating environment, and even the cost, making it extremely challenging for transportation agencies, engineers, and scientists to select the most appropriate solution for their vehicle classification applications. The needs and demands for a comprehensive review of these latest vehicle classification techniques are ever higher.

\Figure[t](topskip=0pt, botskip=0pt, midskip=0pt)[width=.95\textwidth]{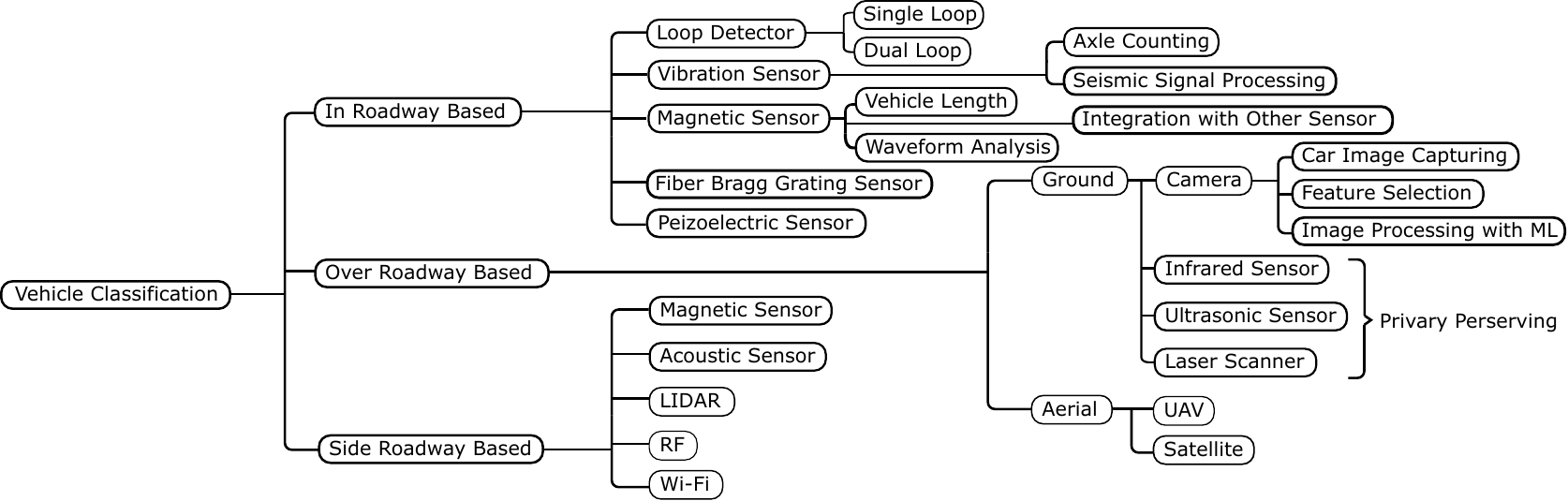}
{The taxonomy of vehicle classification systems.\label{fig:classification_overview}}

In this article, we present a survey on state-of-the-art vehicle classification technologies to address the significant demand and provide guidelines for selecting an appropriate technology for vehicle classification. We systematically organize ideas, research issues, and technical solutions that are developed to achieve high vehicle classification accuracy. Specifically, we classify largely the vehicle classification systems into three categories, \emph{i.e.,} in-road-based, over-road-based, and side-road-based approaches. The vehicle classification schemes in each category is further classified into subcategories based on types of sensors used, methodologies for utilizing the sensors, and mechanisms for classifying vehicles. We provide in-depth description, analysis, and comparison of numerous innovative vehicle classification schemes in each subcategory. We also present a number of open problems and several future research directions.

There are a few surveys on traffic monitoring systems with a focus on vehicle classification. The federal highway administration (FHWA) provides general guidelines for selecting traffic monitoring systems. However, it is limited to industry solutions without discussing on-going research issues and emerging traffic monitoring systems~\cite{2016TrafficGuide}\cite{2016Handbook}. Some papers discuss only traditional traffic monitoring systems such as the loop detectors~\cite{tyburski1988review}. Interestingly, we find that most survey works are focused on vision-based vehicle classification techniques~\cite{datondji2016survey}\cite{tian2011video}\cite{inigo1985traffic}\cite{yousaf2012comparative}\cite{kul2017concise}\cite{jain2019review} overlooking numerous other emerging vehicle classification solutions. Although there are some works that provide a review of vehicle classification systems based on different types of sensors, these papers discuss only a particular type of vehicle classification system such as UAVs~\cite{puri2005survey}\cite{kanistras2015survey}. A comprehensive review on traffic monitoring systems have been performed recently~\cite{bernas2018survey}. However, the paper is concentrated on vehicle detection technologies rather than vehicle classification schemes, which is technically more challenging and has a large body of literature based on emerging technologies. In contrast to existing works, this article provides a comprehensive survey on virtually all vehicle classification technologies developed in the past decade with in-depth analysis of research issues, technical challenges, and novel approaches. The contributions of this article are summarized as follows.

\begin{itemize}
  \item To the best of our knowledge, this article presents the first comprehensive review of latest traffic monitoring systems specifically concentrating on vehicle classification.
  \item This article is specifically focused on discussing various research issues on vehicle classification based on emerging technologies such as machine learning, low-power sensing, networking, and novel image processing algorithms.
   \item This article introduces new breeds of traffic monitoring systems that are significantly different from traditional ones such as RF and Wi-Fi-based traffic monitoring systems.
   \item This article presents open research problems and a number of future research directions.
\end{itemize}

This article is organized as follows. In Section~\ref{sec:vehicle_classification}, the taxonomy of vehicle classification schemes is introduced, followed by detailed descriptions of vehicle classification systems in each category, \emph{i.e.,} in-roadway based systems (Section~\ref{sec:in_roadway}), over-roadway based systems (Section~\ref{sec:over_roadway}), and side-roadway based systems (section~\ref{sec:side_roadway}). We then present open problems and future research directions in Section~\ref{sec:future_research} and conclude in Section~\ref{sec:conclusion}.

\section{Taxonomy of Vehicle Classification Technologies}
\label{sec:vehicle_classification}

This section presents the taxonomy of vehicle classification systems. The details of each vehicle classification system are described in subsequent sections. Vehicle classification systems are largely categorized into three classes depending on where the system is deployed: in-roadway-based, over-roadway-based, and side-roadway-based systems (Fig.~\ref{fig:classification_overview}). We then further classify the vehicle classification systems based on sensor types and how sensor data are analyzed and utilized for vehicle classification.

The in-roadway-based vehicle classification systems install sensors on or under the pavement of a roadway. Different types of sensors are used for the in-roadway-based vehicle classification systems such as piezoelectric sensors~\cite{rajab2014vehicle}, magnetometers~\cite{bottero2013wireless}\cite{xu2018vehicle}, vibration sensors~\cite{stocker2014situational}, loop detectors~\cite{meta2010vehicle}. Various kinds of information is extracted from the sensor data including the vehicle length, axle count, and unique features of the signal/waveform. The in-roadway-based systems boast the high vehicle classification accuracy because the sensors maintain close contact with passing vehicles, effectively capturing the body and motion signature of the vehicles. A major downside is, however, the high cost for installation and maintenance because the pavement of a roadway needs to be sawcut to install the sensors under the roadway. The cost increases significantly due to traffic disruption and lane closure to provide safety to road workers.

The side-roadway-based systems addresses the cost issue of the in-roadway-based vehicle classification schemes since the sensors are installed on a roadside, obviating the needs for lane closure and construction. Similar to the in-roadway-based systems, different types of sensors are adopted for vehicle classification. Some of the most widely used sensors include magnetometers~\cite{wang2014easisee}\cite{yang2015vehicle}, accelerometers~\cite{ma2014wireless}, and acoustic sensors~\cite{george2013vehicle}. Recently, advanced sensors such as Laser Infrared Detection and Ranging (LIDAR)~\cite{lee2012side}\cite{lee2015using}, infrared sensors~\cite{odat2017vehicle}, and Wi-Fi transceivers~\cite{won2017witraffic} have been employed. Despite the benefits of easier installation and reduced cost, the side-roadway-based systems require extra efforts for adjusting precisely the directions and placement of the sensors~\cite{odat2017vehicle}. A more critical problem is that most systems fail to classify overlapped vehicles accurately. Additionally, an algorithm for calibrating the sensor data is needed to mitigate the impact of the noise and increase the classification accuracy.

The over-roadway-based systems utilize sensors installed over the roadway thus being capable of covering multiple lanes simultaneously. For example, unmanned aerial vehicles (UAVs) and satellites are used in these systems~\cite{tang2017arbitrary}. The most prevalent technology under this category is the camera-based systems~\cite{chen2012vehicle}\cite{bautista2016convolutional}. While the camera-based systems have high classification accuracy, the performance is affected by weather and lighting conditions. Another important problem is the driver privacy concerns as there are many people who do not feel comfortable to be exposed to cameras. Some over-roadway-based systems address the privacy concerns by adopting different types of sensors such as infrared sensors~\cite{odat2017vehicle} and laser scanner~\cite{chidlovskii2014vehicle}.

Having presented the taxonomy of the vehicle classification systems which provides a big picture of our vehicle classification schemes, in the following sections, we discuss the details on research issues, technical challenges, hardware/software design, deployment experience, and comparison of various vehicle classification systems.

\section{In-Roadway-Based Vehicle Classification}
\label{sec:in_roadway}

In this section, we discuss various in-road-way-based vehicle classification systems. In reviewing each vehicle classification system, we present the basic theory, specific research problems addressed by the system, and main ideas for vehicle classification. We also discuss vehicle types used for classification and the average vehicle classification accuracy.

Starting with the loop detectors which are the most widely used in-roadway-based vehicle classification systems, we review other in-road-way-based vehicle classification systems built with different types of sensors. The characteristics of the in-roadway-based vehicle classification systems covered in this section are summarized in Table~\ref{table:in-roadway}.

\begin{table*}[htbp]
    \caption{In-roadway-based vehicle classification systems}
    \label{table:in-roadway}
    \center
    \begin{tabular}{| l | l | l | l |l |}
    \hline
    Major Equipment & Publications & Accuracy & Vehicle Classes & Key Features\\ \hline

    \pbox{3cm}{Magnetic Sensors} & \pbox{3cm}{Bottero TRP-C'13~\cite{bottero2013wireless}} & \pbox{1cm}{88\%} & \pbox{3cm}{car, van, truck} & \pbox{6cm}{A wireless sensor network of two magnetic sensors; Vehicle length is used as a key feature} \\ \cline{2-5}

    \pbox{3cm}{} & \pbox{3cm}{Ma TITS'14~\cite{ma2014wireless}} & \pbox{1cm}{99.0\%} & \pbox{3cm}{2,3,4,5,6-axle vehicles} & \pbox{6cm}{Combination of a magnetometor for speed estimation and an accelerometer for axle counting and axle spacing estimation} \\ \cline{2-5}

   \pbox{3cm}{} & \pbox{3cm}{Li Measurement'14~\cite{li2014vehicle}} & \pbox{1cm}{88.9\% (cars), and 94.4\% (busses)} & \pbox{3cm}{cars, and busses} & \pbox{6cm}{A single magnetic sensor; Speed-independent features of vehicle waveform.} \\ \cline{2-5}

    \pbox{3cm}{} & \pbox{3cm}{Li CN'17~\cite{li2017reliable}} & \pbox{1cm}{96.4\%} & \pbox{3cm}{passenger vehicles, SUVs, buses, and Vans} & \pbox{6cm}{Sensor fusion of magnetic waveforms collected from two magnetic sensors that are 80m apart on the same lane} \\ \cline{2-5}

    & \pbox{3cm}{Xu Sensors'18~\cite{xu2018vehicle}} & \pbox{1cm}{95.46\%} & \pbox{3cm}{hatchbacks, sedans, buses, and multi-purpose vehicles} & \pbox{6cm}{Advanced machine learning techniques for classification focusing on the imbalance effect} \\ \cline{2-5}

    & \pbox{3cm}{Balid TITS'18~\cite{balid2018intelligent}} & \pbox{1cm}{97\%} & \pbox{3cm}{passenger vehicles, single-unit trucks, combination trucks, and multi-trailer trucks.} & \pbox{6cm}{Machine learning-based classification using the vehicle length as a key feature} \\ \cline{2-5}

    & \pbox{3cm}{Dong Access'18~\cite{dong2018improved}} & \pbox{1cm}{80.5\%} & \pbox{3cm}{class 1 (sedans and SUVs), class 2 (vans and seven-seat cars), class 3 (light and medium trucks), and class 4 (heavy trucks and semi trailers) - total of 2,231 vehicles} & \pbox{6cm}{Classification based on XGBoost using a single magnetic sensor} \\ \hline

    \pbox{3cm}{Vibration Sensors} & \pbox{3cm}{Bajwa IPSN'11~\cite{bajwa2011pavement}} & \pbox{1cm}{N/A} & \pbox{3cm}{vehicles with different axle counts and spacing (mostly large trucks)} & \pbox{6cm}{Axle count and spacing between axles as key features} \\ \cline{2-5}

    & \pbox{3cm}{Stocker TITS'14~\cite{stocker2014situational}} & \pbox{1cm}{83\%} & \pbox{3cm}{light, and heavy vehicles} & \pbox{6cm}{Unique characteristics of seismic signals used as key features; Multilayer perceptron (MLP) feedforward artificial neural networks for classification} \\ \cline{2-5}

    & \pbox{3cm}{Zhao TRR'18~\cite{zhao2018vibration}} & \pbox{1cm}{89.4\%} & \pbox{3cm}{passenger car, bus, and 2$\sim$6 axle trucks/trailers} & \pbox{6cm}{Axle count and spacing between axles as key features; Capable of classifying 2-axle cars with similar axle configurations based on a multi-parameter classifier} \\ \cline{2-5}

    & \pbox{3cm}{Jin GRSL'18~\cite{jin2018vehicle}} & \pbox{1cm}{92\%} & \pbox{3cm}{assault Amphibian Vehicle (AAV) and dragon wagon (DW)} & \pbox{6cm}{Focused on the complexity of seismic signal; Convolutional neural network (CNN) with the log-scaled frequency cepstral coefficient (LFCC) matrix as a key feature to address the complexity} \\ \hline

    \pbox{3cm}{Loop detectors} & \pbox{3cm}{Meta TVT'10~\cite{meta2010vehicle}} & \pbox{1cm}{94.2\%} & \pbox{3cm}{car/jeep, minibus/van, pickup/truck, bus, and motorcycle} & \pbox{6cm}{Noise reduction in the raw signal; PCA for dimensionality reduction; Application of BPNN} \\ \cline{2-5}

   \pbox{3cm}{} & \pbox{3cm}{Tok TRB'10~\cite{tok2010vector}} & \pbox{1cm}{80.8\%} & \pbox{3cm}{axle configuration classes (27), drive unit body classes (9), and trailer unit body classes (10)} & \pbox{6cm}{Combination of axle configuration-based and inductive signature-based systems} \\ \cline{2-5}


   \pbox{3cm}{} & \pbox{3cm}{Jeng TRR'13~\cite{jeng2013wavelet}} & \pbox{1cm}{93.8\%} & \pbox{3cm}{The 13 FHWA vehicle classes~\cite{FHWAClass}} & \pbox{6cm}{Feature extraction from inductive signals using the wavelet transformation technique; Classification with k-NN} \\ \cline{2-5}

   & \pbox{3cm}{Lamas Sensors'15~\cite{lamas2015vehicle}} & \pbox{1cm}{96\%} & \pbox{3cm}{car, truck, and van} & \pbox{6cm}{Spectral features of inductive signatures using DFT} \\ \cline{2-5}

    & \pbox{3cm}{Liu TRP-C'14~\cite{liu2014length}} & \pbox{1cm}{99.4\%} & \pbox{3cm}{long vehicles, regular cars (total of 2,547 cars)} & \pbox{6cm}{Vehicle-length-based work using a single loop detector; A traffic theory was used to estimate vehicle speed; Classifies vehicles into only two types} \\ \cline{2-5}

    & \pbox{3cm}{Wu TRR'14~\cite{wu2014vehicle}, TRP-C'14~\cite{wu2014improved}} & \pbox{1cm}{99\%} & \pbox{3cm}{three length classes with boundaries at 28 ft and 46 ft} & \pbox{6cm}{Addresses the issue of non-zero acceleration of passing cars } \\ \hline

    \pbox{3cm}{Weigh In Motion} & \pbox{3cm}{Hernandez TRP-C'16~\cite{hernandez2016integration}} & \pbox{1cm}{+80.0\%} & \pbox{3cm}{31 single and semi-trailer body trucks, and 23 single unit trucks (total of 18,967 trucks)} & \pbox{6cm}{Classification for truck body types; Integration of WIM with an inductive loop detector} \\ \hline

   \pbox{3cm}{Piezoelectric Sensors} & \pbox{3cm}{Rajab IV'14~\cite{rajab2014vehicle}} & \pbox{1cm}{86.9\%} & \pbox{3cm}{The 13 FHWA vehicle classes~\cite{FHWAClass}} & \pbox{6cm}{An array of piezoelectric sensors; Standard features are used} \\ \hline

  \pbox{3cm}{Fiber Bragg Grating Sensors} & \pbox{3cm}{Huang IEEE Sensors'18~\cite{huang2018vehicle}} & \pbox{1cm}{98.5\%} & \pbox{3cm}{small, medium, large} & \pbox{6cm}{A sensor network of FBG sensors; Standard features are used} \\ \hline

    \end{tabular}
\end{table*}

\subsection{Loop Detectors}
\label{sec:over_loop_detectors}

\Figure[t](topskip=0pt, botskip=0pt, midskip=0pt)[width=.95\columnwidth]{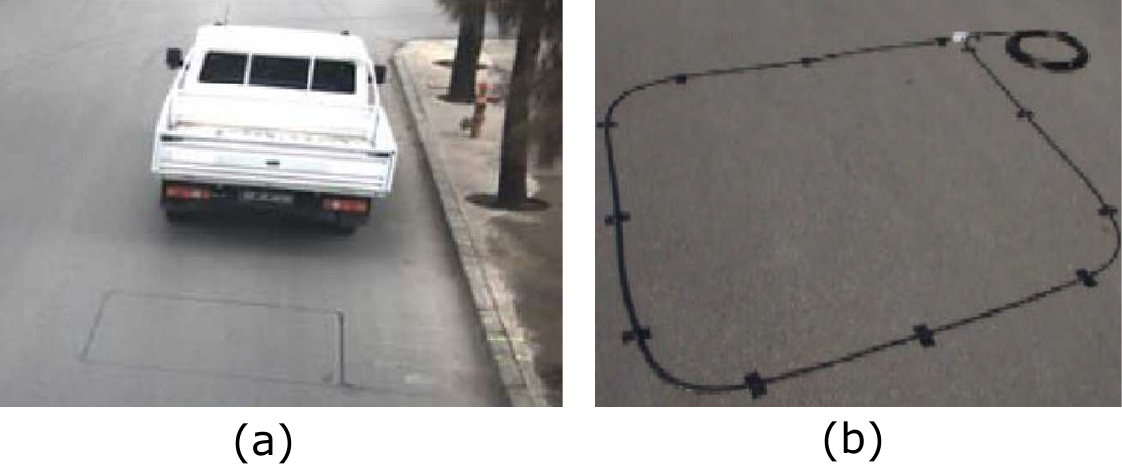}
{Types of loop detectors: (a) saw-cut loop~\cite{meta2010vehicle}; (b) preformed loop~\cite{martin2003detector}.\label{fig:loop_example}}

An inductive loop detector is one of the most commonly used traffic monitoring systems for vehicle detection and classification~\cite{coifman2014improved}. It is a coil of wire that is embedded under the road surface (Fig.~\ref{fig:loop_example}). It captures the change of inductance and generates a time-variable signal when a vehicle passes over. The characteristics of the signal such as the amplitude, phase, and frequency spectrum are varied depending on the classes of vehicles. These unique characteristics of the signal are known as the magnetic profile~\cite{jeng2014high}, which is used to perform vehicle classification.

There are largely two types of loop detectors depending on the installation method: saw-cut and preformed loops. The saw-cut method requires to saw-cut the pavement, lay the loop wire, and protect the wire by filling the pavement (Fig.~\ref{fig:loop_example}(a)). The preformed loop detectors do not embed the loop wire under the pavement; instead it encases the loop wire in a PVC pipe and attach the pipe on the pavement (Fig.~\ref{fig:loop_example}(b)). The loop detectors can also be categorized into the single loop detectors and dual loop detectors depending on the number of loop detectors used for vehicle classification. The dual loop detectors consist of a pair of loop detectors in a lane. A key strength of the dual loop detectors compared with the single loop detectors is that the dual loop detectors can measure the vehicle speed and vehicle length based on the predetermined longitudinal distance between the two loop detectors.

Numerous research works have been conducted to enhance the performance of loop detectors for vehicle classification. In particular, recent development of machine learning technologies sparked the emergence of advanced loop detectors that apply machine learning techniques to analyze the magnetic signature of passing cars. Meta and Cinsdikici adopt the backpropagation neural network (BPNN) for vehicle classification~\cite{meta2010vehicle}. Specifically, based on the observation that the low classification accuracy of existing loop detectors is attributed to simple data sampling of noisy raw signals, an algorithm based on Discrete Fourier transform (DFT) is designed to clear the noise. The principal Component Analysis (PCA) is then applied to reduce the dimensionality of the noiseless data. The PCA features are expanded to emphasize the undercarriage height variation of a passing vehicle. Finally, the output of PCA is fed into the three-layered BPNN to classify the vehicles into five classes: car/jeep, minibus/van, pickup/truck, bus, and motorcycle. The average classification accuracy of 94.2\% was achieved.

A significant technical challenge for classifying vehicles using loop detectors is that vehicles with similar axle configurations are difficult to classify accurately. Tok and Ritchie address this challenge by developing a vehicle classification system that effectively combines the axle-based vehicle classification method with the vehicle body signature-based classification method~\cite{tok2010vector}. More specifically, vehicles are first classified into three high-level types based on the number of axle clusters. And then, the vehicle body signature (\emph{i.e.,} the magnetic profile of the passing vehicle) is used to further classify the vehicles based on the multi-layer feedforward neural network (MLF)~\cite{svozil1997introduction}. The proposed system achieved 80.8\% vehicle classification accuracy for a total of 1,029 vehicles with 27 different axle configurations, 9 drive unit body classes, and 10 trailer unit body classes. It can be noted that the seemingly low classification accuracy of 80.8\% is actually high considering the large number of vehicles with similar axle configurations as well as the large number of vehicle types used for performance evaluation.


Jeng \emph{et al.} develop a similar vehicle classification system based on the analysis of the magnetic signature of a passing car~\cite{jeng2013wavelet}. Their contribution is that the Haar wavelet transformation technique~\cite{walnut2013introduction} is adopted to compress the waveform data, thereby removing the salient characteristics of vehicle signatures and to maintain more distinctive features in the compressed data. After compressing the waveform data, the \emph{k}-nearest neighbor (\emph{k}NN) method is used as a classifier to classify vehicles into 13 FHWA vehicle types~\cite{FHWAClass}. A data set collected from the I-405 of the city of Irvine, CA, as well as the data set obtained from the city of San Onofre, CA were used for evaluating the performance of the classification system. The average classification accuracy was 93.8\%.

The vehicle classification systems discussed thus far are based on a single loop detector. However, a limitation of the single loop detector is that it does not allow to measure the vehicle speed which can be used to derive the vehicle body length, as apposed to the dual loop detector. Specifically, the dual loop detector consists of a pair of loop detectors in a lane~\cite{cheevarunothai2006identification} and measures the traversal time of a passing vehicle, which is used to calculate the vehicle speed by  dividing the traversal time by the known distance between the pair of the two loop detectors. The body length of a passing vehicle can then be calculated by multiplying the speed with the dwell time over a loop detector.

Taking the advantage of dual loop detectors, numerous vehicle classification systems are developed. A notable aspect of these systems is that the vehicle length is used as a key feature for vehicle classification ~\cite{cheevarunothai2006identification}. In particular, Wu~\emph{et al.} note that small changes in acceleration may influence the precision of estimating the vehicle length, consequently degrading the classification accuracy significantly especially under congested conditions~\cite{wu2014vehicle}\cite{wu2014improved}. To address this challenge, they develop a new method that takes into account the possibility of the non-zero acceleration of a passing vehicle. The new approach was evaluated using the Next Generation Simulation (NGSIM) datasets~\cite{kovvali2007video}. Specifically, vehicles were classified based on their lengths, \emph{i.e.,} with boundaries at  28 ft and 46 ft. The average classification accuracy was over 98\%. While dual loop detectors may yield better results by using the vehicle length as an additional feature for vehicle classification, classifying vehicles with similar body lengths (\emph{e.g.,} pick-up trucks and minivans) still remains as a challenge.

Despite the better classification results that dual loop detectors provide, a major disadvantage of these classification systems is the high cost compared to single loop detectors. Fortunately, researchers recently develop advanced schemes that perform vehicle classification accurately even with a single loop detector. For example, Lamas-Seco \emph{et al.} identify that certain spectral features extracted from the magnetic signal collected from a signal loop detector have no dependency with the vehicle speed~\cite{lamas2015vehicle}. Specifically, they argue that based on these features, an effective classification system can be developed without relying on dual loop detectors. They classified vehicles into three types: car, truck, and van. The average classification accuracy was about 96\%.

Liu and Sun also address the limitation of single loop detectors, successfully measuring the vehicle length with a single loop detector and using it as a key feature for vehicle classification~\cite{liu2014length}. Newell's simplified car following model~\cite{newell2002simplified} is adopted to understand the relationships among vehicles in a platoon and estimate the vehicle occupation time. The classification is performed by comparing the anticipated vehicle occupation time with the measured vehicle occupation time, where the discrepancy indicates a long vehicle. Field data collected from a highway with a total of 2,547 samples were used for the experiments. The average classification accuracy was 99.4\%.


\subsection{Magnetic Sensors}
\label{sec:sub_magnetic}

\Figure[t](topskip=0pt, botskip=0pt, midskip=0pt)[width=.95\columnwidth]{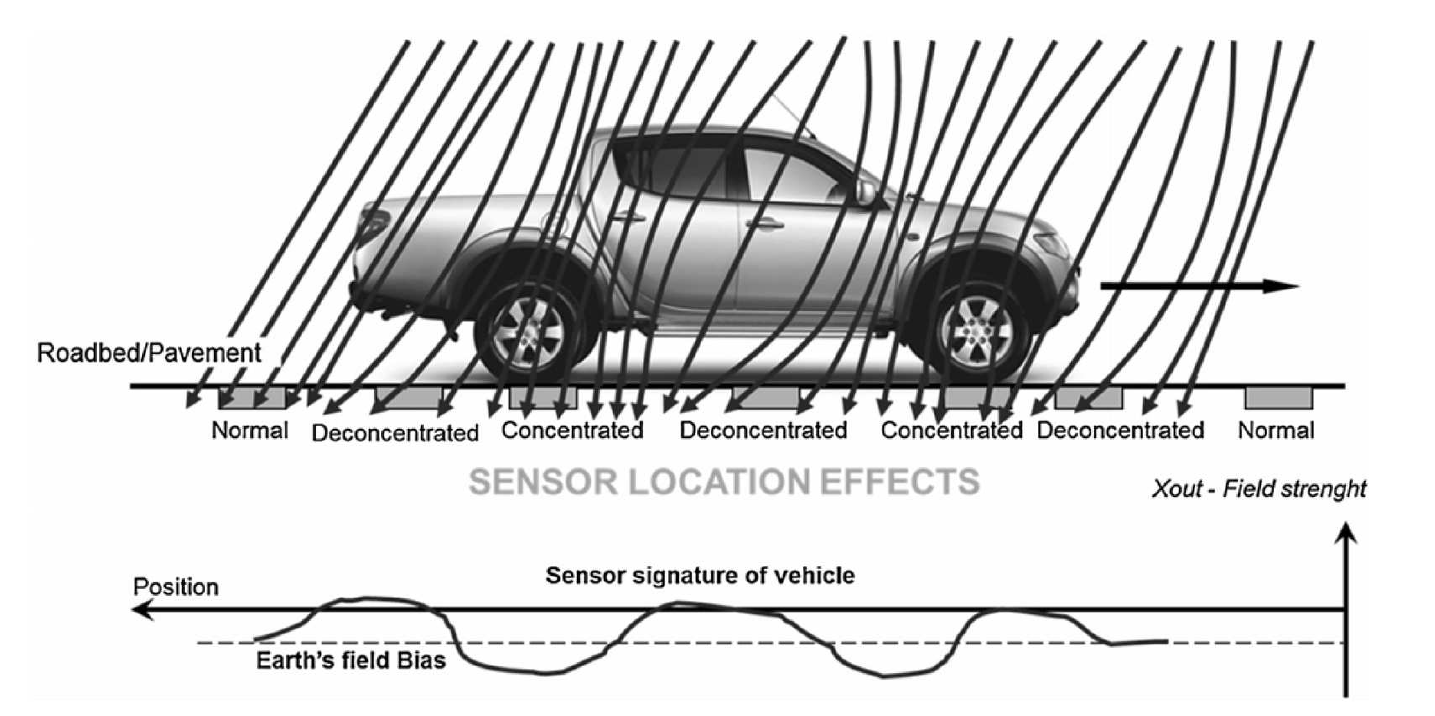}
{Magnetic field changes by a vehicle~\cite{bottero2013wireless}.\label{fig:mag}}

A large amount of ferrous metals in a vehicle frame induces disturbance to the Earth's magnetic field~\cite{cheung2005traffic}. Fig.~\ref{fig:mag} illustrates the distortion of the magnetic field caused by a passing vehicle. Magnetic sensors are used to classify vehicles by capturing the distinctive changes in the magnetic field which depend on different vehicle body types. Compared to loop detectors, magnetic sensor-based vehicle classification systems have advantages in terms of the size, weight, cost, and energy efficiency. In this section, we review recent research works for developing vehicle classification systems based on magnetic sensors.



We categorize magnetic sensor-based vehicle classification systems into three types: (1) a system consisting of multiple magnetic sensors networked through a wireless sensor network -- a primary classification for such systems is the vehicle length; (2) a system based on a single magnetic sensor -- these systems rely on the magnetic waveform analysis mostly using machine learning techniques; (3) a hybrid system that exploits both the magnetic waveform analysis and the vehicle length information obtained through a wireless sensor network of magnetic sensors.

Bottero \emph{et al.} create a wireless sensor network (WSN) consisting of two magnetic sensors to perform vehicle classification~\cite{bottero2013wireless}. Specifically, two pavement-mounted magnetic sensors are aligned to the lane axis to measure the vehicle speed. Given the distance between the two sensors, the vehicle length can be calculated. Vehicle classification is then performed using the vehicle length as a key feature, which is similar to the vehicle classification method of dual loop detectors~\cite{cheevarunothai2006identification}. In this classification system, vehicles are classified into three different types, namely cars, vans, and trucks and the average classification accuracy of 88\% was achieved.

Balid \emph{et al.} propose a similar approach as Bottero \emph{et al.}~\cite{bottero2013wireless} that uses the vehicle length as a key feature for performing vehicle classification~\cite{balid2018intelligent}. In particular, the main feature is called the vehicle magnetic length which is defined as the product of the vehicle speed and the period of time that the vehicle was on a magnetic sensor. Specifically, the vehicle speed is measured by calculating the travel time between two longitudinally located magnetic sensors with known distance between them. Using the vehicle magnetic length as the main feature, different machine learning classifiers are adopted for vehicle classification such as Decision Tree (DT), support vector machine (SVM), k-Nearest Neighbor (\emph{k}NN), and Naive Bayes Classifier (NBC). The classification accuracy was over 97\% for classifying vehicles into passenger vehicles, single-unit trucks, combination trucks, and multi-trailer trucks.

Li and Lv propose another wireless sensor network consisting of magnetic sensors for vehicle classification~\cite{li2017reliable}. Similar to~\cite{bottero2013wireless}, two magnetic sensors are deployed on the same lane that are 80m apart from each other. Unlike other works, the proposed vehicle classification system utilizes magnetic sensors not only for deriving the vehicle length but also for obtaining magnetic waveforms to perform data analysis to enhance the classification accuracy. Specifically, the main contributions of their work compared with other solutions based on a wireless sensor network of magnetic sensors are two fold. First, a novel data segmentation technique is developed to separate the magnetic waveform effectively from the overall waveform of magnetic sensor data. Second, a sensor fusion algorithm is developed to correlate the feature waveforms from the two magnetic sensors to improve the classification accuracy. Consequently, the average classification accuracy was 96.4\% in classifying vehicles into four types: passenger vehicles, SUVs, busses, and vans.

Different types of sensors have been integrated with magnetic sensors to enhance the effectiveness of vehicle classification. For example, Ma \emph{et al.} propose a WSN consisting of magnetic sensors and accelerometers~\cite{ma2014wireless}. Specifically, magnetic waveforms collected from deployed magnetic sensors are used to estimate the vehicle speed, and the accelerometer is used to count the number of axles and estimate the axle spacing between each pair of axles by using the vehicle speed information. Vehicles are classified according to the FHWA 13-category~\cite{wyman1985field}. The proposed system classifies vehicles with the accuracy of 99\%. While the resulting classification accuracy seems very high, it is challenging to sustain such high accuracy for vehicles with the same axle count and similar axle spacing.

Recent research shows that it is possible to achieve high classification accuracy using only a single magnetic sensor with the help of advanced machine learning techniques. The idea is to leverage machine learning techniques to automatically extract useful features from magnetic waveforms rather than relying on simple features such as the peaks of waveforms and to build a vehicle classification model effectively for vehicle classification~\cite{xu2018vehicle}. Various machine learning techniques are used for vehicle classification such as the k-nearest neighbor (\emph{k}NN)~\cite{keller1985fuzzy}, support vector machine (SVM)~\cite{suykens1999least}, back-propagation neural network (BPNN)~\cite{goh1995back}, and convolutional neural network (CNN)~\cite{krizhevsky2012imagenet}.

Li \emph{et al.} identifies eight speed-independent features (\emph{i.e.,} number of peaks, the maximum peak time ratio, the minimum trough time ratio, the mean value, the standard deviation, the maximum peak amplitude, the minimum trough amplitude, and the maximum peak/trough amplitude ratio) from a magnetic waveform~\cite{li2014vehicle}. These features are then used to build a vehicle classification model based on the optimal Minimum Number of Split-sample (MNS)-based Classification and
Regression Tree (CART) algorithm~\cite{steinberg2009cart}. They achieve the classification accuracy of 88.9\%, and 94.4\% for cars and busses, respectively. Especially, Xu \emph{et al.} focus on the problem of the unbalanced magnetic sensor dataset~\cite{xu2018vehicle}. They note that the number of vehicles in each vehicle class is significantly different from each other in many open datasets which often leads to degraded classification performance. The proposed work thus aims to minimize the imbalance effect and applies the \emph{k}-nearest neighbor (\emph{k}NN) for vehicle classification.


Dong \emph{et al.} also show that a single magnetic sensor can be a powerful tool for vehicle classification~\cite{dong2018improved}. Three types of features are extracted from the Z-axis of a magnetic signal including statistical, energy, and short-term features. In particular, the energy features are used because it is highly correlated with the vehicle size. These features are provided as input to a classifier, XGBoost~\cite{chen2016xgboost} to perform vehicle classification into four categories: class 1 (sedans and SUVs), class 2 (vans and seven-seat cars), class 3 (light and medium trucks), and class 4 (heavy trucks and semi trailers). The average classification accuracy was 80.5\% with 1,797 vehicles being successfully classified out of 2,231 vehicles.

\subsection{Vibration Sensors}
\label{sec:sub_vibration}

Using highway pavement itself as a transducer, vibration sensors capture the unique vibration patterns induced by passing vehicles due to the low elasticity of road pavement that makes vibrations well localized in time and space~\cite{bajwa2011pavement}. Fig.~\ref{fig:vibration} shows an example of a vibration sensor and the installation process. Researchers address numerous challenges of vibration sensor-based classification systems such as the effect of the underlying geology on the propagation of the seismic wave, and the complex nature of seismic waveforms in terms of the forms, directions, and speeds, which make vehicle classification based on vibration sensor data very difficult.

\Figure[t](topskip=0pt, botskip=0pt, midskip=0pt)[width=.95\columnwidth]{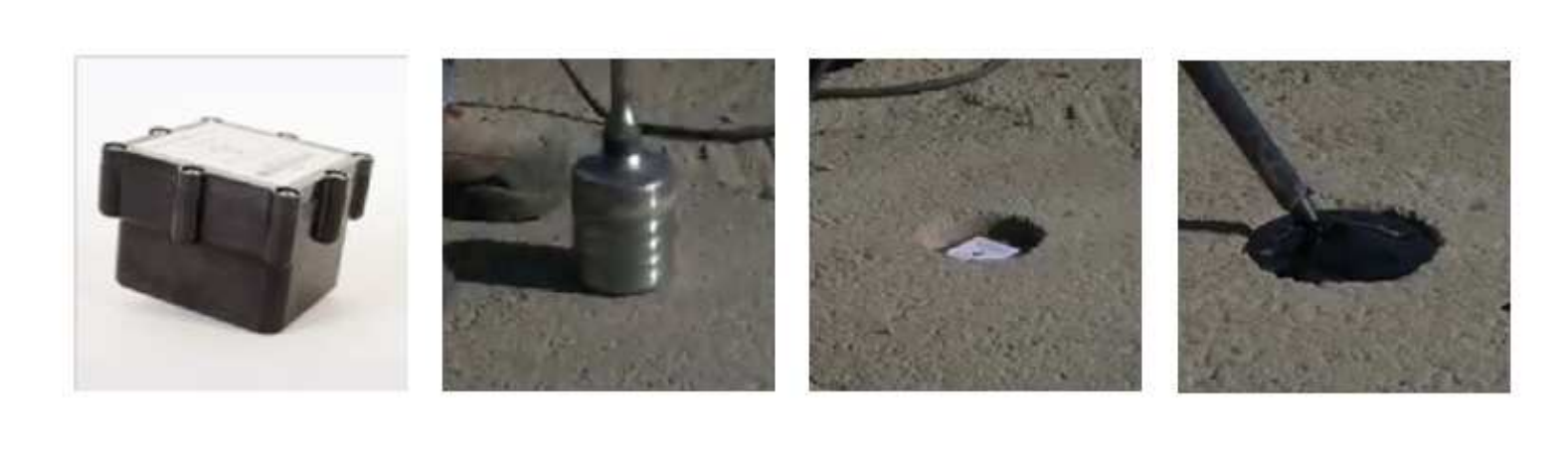}
{An example of a vibration sensor~\cite{bajwa2011pavement}.\label{fig:vibration}}

Various vibration sensor-based vehicle classification systems are developed to address the aforementioned challenges. These systems can be divided largely into two types: the classification systems that utilize vibrations to count the number of axles and measure the spacing between axles and use the axle count and spacing as main features for vehicle classification~\cite{bajwa2011pavement}, and the systems based on the analysis of seismic waveforms induced by passing vehicles for vehicle classification~\cite{stocker2014situational}\cite{jin2018vehicle}, \emph{i.e.,} the characteristic features of the seismic waves are extracted to model a classifier to perform classification; Since the seismic waves are very complex, machine learning techniques are often adopted to extract features effectively.

Bajwa \emph{et al.} propose a vehicle classification system based on the axle count and spacing between axles~\cite{bajwa2011pavement}. The proposed system consists of magnetic sensors and vibration sensors. The magnetic sensors are used for detecting a vehicle and reporting the arrival and departure times of the passing vehicles. The vibration sensors are utilized for calculating the number of axles and spacing between axles which are the two key features for vehicle classification.


Zhao \emph{et al.} develop a novel vibration sensing system for vehicle classification called the distributed optical vibration sensing system (DOVS)~\cite{zhao2018vibration}. While DOVS is similar to~\cite{bajwa2011pavement} in that the axle count and the spacing between axles are used as main features, the main contribution is that DOVS achieves higher resistance to damage and electromagnetic interference, making DOVS more reliable in severe environments. Furthermore, it is easy to deploy and the cost for installation is low compared with other vibration sensor-based systems. Another notable feature of DOVS is the support for classification of the vehicles with similar axle configurations especially 2-axle vehicles such as vans, two-axle buses, and two-axle trucks by developing a multi-parameter classifier incorporating additional features in the frequency domain and the vehicle speed. DOVS classifies vehicles into 10 vehicle types and achieves the average classification accuracy of 89.4\%.

Other vibration sensor-based vehicle classification systems exploit the unique characteristics of the seismic signals induced by passing vehicles. Stocker \emph{et al.}~\cite{stocker2014situational} propose a digital signal processing algorithm to process vibration sensor data to identify unique vibration patterns for passing vehicles. In particular, a machine learning technique is applied to collected vibration sensor data to perform vehicle classification. Specifically, the multilayer perceptron (MLP) feedforward
artificial neural networks~\cite{haykin1994neural} is adopted to classify vehicles into light (mini-cleaner, mini-lifter, personal-car, van, ambulance, fire-van, and pickup-truck) and heavy vehicles (truck, fire-truck, and bucket-digger). The classification accuracy obtained was 83\%.

A similar work based on the analysis of seismic waveforms is performed by Jin \emph{et al.}~\cite{jin2018vehicle}. The authors focus on the complexity of the seismic signals which is nonstationary and nonlinear. The seismic signal comprises a number of signals generated by a passing vehicle (\emph{e.g.,} the engine and propulsion system of a passing car). It is not only highly dependent on underlying geology, but its propagation speed and direction vary significantly~\cite{jin2012target}. To achieve high classification accuracy under the complexity of the seismic signal, the authors apply a convolutional neural nework (CNN). Specifically, they develop a seismic signal-based deep CNN architecture for classifying vehicles. The proposed CNN framework takes the log-scaled frequency cepstral coefficient (LFCC) matrix as a key feature. Vehicle classification was performed with the vibration sensor data collected from the DARPA’s SensIt project for two vehicle classes, \emph{i.e.,} Assault Amphibian Vehicle (AAV) and dragon wagon (DW). The best classification accuracy achieved was 92\%.

\subsection{Other Technologies}
\label{in_road_other_tech}

Various kinds of sensors such as weigh-in-motion sensors~\cite{hernandez2016integration}, peizoelectric sensors~\cite{rajab2014vehicle}, and fiber-optic sensors~\cite{huang2018vehicle} are used to develop in-roadway-based vehicle classification systems. It is interesting to note that while these sensors are less frequently used for vehicle classification compared to loop detectors, vibration sensors, and magnetic sensors, similar features such as the axle count, axle spacing, and vehicle length are used for vehicle classification.

Hernandez \emph{et al.} develop an in-road-based vehicle classification system that integrates a weigh-in-motion sensor with a loop detector~\cite{hernandez2016integration}. Specifically, the vehicle weight data are combined with the axle spacing data to achieve better classification accuracy. They propose to utilize multiple classification models, \emph{i.e.,} Naive Bayes Classifier, Decision Tree, SVM, Multilayer Feed forward Neural Network~\cite{hornik1989multilayer}, and Probabilistic Neural Network~\cite{specht1990probabilistic}. In particular, a multiple classifier systems (MCS) method~\cite{kuncheva2004combining} is adopted to combine the results of these classifiers. A huge dataset of 18,967 trucks was used to classify the trucks into 31 single and semi-trailer body trucks, and 23 single unit trucks. The accuracy was over 80\% for each truck body type.


Rajab \emph{et al.} develop a multi-element piezoelectric sensor system which consists of 16 piezoelectric sensors~\cite{rajab2014vehicle}. Three main features, \emph{i.e.,} the number of tires, vehicle length, and axle spacing are used for vehicle classification. Specifically, the proposed system estimates the number of tires by sensing impact on multiple sensor elements. Additionally, the vehicle speed is estimated based on the time difference between impacts on two sensors aligned to the lane axis. The vehicle length and the axle spacing are computed based on the vehicle speed and the dwell time over a sensor. The 13 FHWA vehicle classes were used for vehicle classification. The average classification accuracy was 86.9\%.

Recent advances in fiber-optic sensors that are small, lightweight, immune to electro-magnetic interference gave rise to novel traffic engineering applications~\cite{malla2008special}. Huang \emph{et al.} adopt fiber bragg grating (FBG) sensors for vehicle classification~\cite{huang2018vehicle}. A sensor network consisting of two FBG sensors is developed to extract the features of the number of axles and axle spacing. Specifically, the FBG sensors capture the strain signals generated from the pavement when vehicles pass on the road, so that an individual peak is used to identify the features. With the two aligned sensors, the vehicle speed can be measured, and the axle spacing is measured based on the vehicle speed. The classification accuracy was high as 98.5\% partly due to the simple vehicle classification scheme, \emph{i.e.,} small, medium, and large vehicles.

\section{Over-Roadway-Based Vehicle Classification}
\label{sec:over_roadway}

\begin{table*}[htbp]
\caption{Over-roadway-based vehicle classification systems}
\label{table:over_roadway}
\center
    \begin{tabular}{| l | l | l | l |l |}
    \hline
    Major Equipment & Publications & Accuracy & Vehicle Classes & Key Features \\ \hline

   \pbox{3cm}{Camera} & \pbox{3cm}{Chen ITSC'12~\cite{chen2012vehicle}} & \pbox{1cm}{94.6\%} & \pbox{3cm}{motorcycles, cars, vans, buses, and unknown vehicles} & \pbox{6cm}{GMM for background noise removal; SVM for classification} \\ \cline{2-5}

   \pbox{3cm}{} & \pbox{3cm}{Mithun TITS'12~\cite{mithun2012detection}} & \pbox{1cm}{+88\%} & \pbox{3cm}{motorbikes, rickshaws, autorickshaws, cars, jeeps, covered vans, and busses} & \pbox{6cm}{Vehicle detection based on multiple virtual detection lines (MVDLs); A two-step classification method based on \emph{k}NN} \\ \cline{2-5}

  \pbox{3cm}{} & \pbox{3cm}{Unzueta TITS'12~\cite{unzueta2012adaptive}} & \pbox{1cm}{92.6\%} & \pbox{3cm}{two wheels, light vehicles, and heavy vehicles} & \pbox{6cm}{Addressed the problem of the dynamic changes of the background; A multicue background subtraction method} \\ \cline{2-5}

   \pbox{3cm}{} & \pbox{3cm}{Dong TITS'15~\cite{dong2015vehicle}} & \pbox{1cm}{89.4\%} & \pbox{3cm}{truck, minivan, bus, passenger car, and sedan} & \pbox{6cm}{A two-stage CNN for automatic feature extraction; Softmax classifier based on multi-task learning} \\ \cline{2-5}

   \pbox{3cm}{} & \pbox{3cm}{Karaimer ITSC'15~\cite{karaimer2015combining}} & \pbox{1cm}{96.5\%} & \pbox{3cm}{cars, vans, and motorcycles} & \pbox{6cm}{Combination of \emph{k}NN with shape-based features and SVM with HOG features} \\ \cline{2-5}

   \pbox{3cm}{} & \pbox{3cm}{Huttunen IV'16~\cite{huttunen2016car}} & \pbox{1cm}{97.0\%} & \pbox{3cm}{bus, truck, van and small car (total of 6,555 cars)} & \pbox{6cm}{Automatically extracted features using DNN} \\ \cline{2-5}

   \pbox{3cm}{} & \pbox{3cm}{Adu-Gyamfi TRR'17~\cite{adu2017automated}} & \pbox{1cm}{+89\%} & \pbox{3cm}{The 13 FHWA vehicle classes~\cite{FHWAClass}} & \pbox{6cm}{Deep convolutional neural network for feature extraction and SVM for classification. Pretraining DCNN model with auxiliary data} \\ \cline{2-5}

   \pbox{3cm}{} & \pbox{3cm}{Javadi PCS'17~\cite{javadi2018vehicle}} & \pbox{1cm}{96.5\%} & \pbox{3cm}{private cars, light trailers, buses, and heavy trailers} & \pbox{6cm}{Designed for classifying vehicles with similar body dimensions; Prior knowledge about speed regulations used for enhanced performance} \\ \cline{2-5}

   \pbox{3cm}{} & \pbox{3cm}{Zhao TCDS'17~\cite{zhao2017deep}} & \pbox{1cm}{97.9\%} & \pbox{3cm}{sedans, vans, trucks, SUVs, and coaches} & \pbox{6cm}{The visual attention mechanism to focus on only relevant part of the car image} \\ \cline{2-5}

   \pbox{3cm}{} & \pbox{3cm}{Thegarajan CVPRW'17~\cite{theagarajan2017eden}} & \pbox{1cm}{97.8\%} & \pbox{3cm}{articulated trucks, background, busses, bicycles, cars, motorcycles, nonmotorized vehicles, pedestrians, pickup trucks, single unit trucks, and work vans (total of 786,702 cars)} & \pbox{6cm}{Used the largest image dataset ever known to the research community} \\ \cline{2-5}

  \pbox{3cm}{} & \pbox{3cm}{Kim CVPRW'17~\cite{kim2017vehicle}} & \pbox{1cm}{97.8\%} & \pbox{3cm}{articulated trucks, background, busses, bicycles, cars, motorcycles, nonmotorized vehicles, pedestrians, pickup trucks, single unit trucks, and work vans} & \pbox{6cm}{Data augmentation; A weighting scheme to compensate for different sample sizes} \\ \cline{2-5}

  \pbox{3cm}{} & \pbox{3cm}{Liu ACCESS'17~\cite{liu2017ensemble}} & \pbox{1cm}{97.6\%} & \pbox{3cm}{articulated truck, background, bicycle, bus, car, motorcycle, pedestrian, pickup truck, non-motorized vehicle, single unit truck and work van} & \pbox{6cm}{Data augmentation; An ensemble of CNN models} \\ \cline{2-5}

  \pbox{3cm}{} & \pbox{3cm}{Chang ITSM'18~\cite{chang2018vision}} & \pbox{1cm}{97.6\%} & \pbox{3cm}{sedans, SUVs, vans, busses, and
  trucks (total of 136,726 cars)} & \pbox{6cm}{The multi-vehicle occlusion problem was addressed prior to vehicle classification} \\ \cline{2-5}

  \pbox{3cm}{} & \pbox{3cm}{Hasnat ICIP'18~\cite{hasnat2018new}} & \pbox{1cm}{99.0\%} & \pbox{3cm}{light, intermediate, heavy, heavy with more than 2 axles, and motorbikes} & \pbox{6cm}{A camera integrated with optical sensors; Two combined classifiers using the Gradient Boosting technique} \\ \hline

   \pbox{3cm}{Aerial Platforms} & \pbox{3cm}{Cao ICIP'11~\cite{cao2011linear}} & \pbox{1cm}{90.0\%} & \pbox{3cm}{only for vehicle detection} & \pbox{6cm}{Not capable of vehicle classification} \\ \cline{2-5}

   \pbox{3cm}{} & \pbox{3cm}{Liu GRSL'15~\cite{liu2015fast}} & \pbox{1cm}{Up to 98.2\%} & \pbox{3cm}{cars, and trucks} & \pbox{6cm}{HOG features used with a single hidden layer neural network} \\ \cline{2-5}

  \pbox{3cm}{} & \pbox{3cm}{Audebert RS'17~\cite{audebert2017segment}} & \pbox{1cm}{80.0\%} & \pbox{3cm}{sedans, vans, pickups, trucks} & \pbox{6cm}{Data normalization and augmentation schemes to reduce discrepancy between training and testing datasets; LeNet, AlexNet, and VGG-16 for vehicle classification} \\ \cline{2-5}

  \pbox{3cm}{} & \pbox{3cm}{Tan ICIP'18~\cite{tan2018vehicle}} & \pbox{1cm}{80.3\%} & \pbox{3cm}{sedans, vans, pickups, trucks} & \pbox{6cm}{Manned aerial vehicle equipped with an infrared sensor; AlexNet and Inception Model for classification} \\ \hline

    \pbox{3cm}{Infrared + ultrasonic sensors} & \pbox{3cm}{Odat TITS17~\cite{odat2017vehicle}} & \pbox{1cm}{Up to 99\%} & \pbox{3cm}{sedan, pickup truck, SUV, bus, two wheeler} & \pbox{6cm}{Combination of infrared sensors and ultrasonic sensors; Classification based on the Bayesian Network and Neural Network} \\ \hline

    \pbox{3cm}{Laser scanner} & \pbox{3cm}{Sandhawalia ITSC'13~\cite{sandhawalia2013vehicle}} & \pbox{1cm}{82.5\%} & \pbox{3cm}{passenger vehicles, passenger vehicles with one trailer, trucks, trucks with one trailer, trucks with two trailers, and motorcycles} & \pbox{6cm}{Representation of a laser scanner profile as an image} \\ \cline{2-5}

    \pbox{3cm}{} & \pbox{3cm}{Chidlovskii ITSC'14~\cite{chidlovskii2014vehicle}} & \pbox{1cm}{86.8\%} & \pbox{3cm}{passenger vehicles, passenger vehicles with one trailer, trucks, trucks with one trailer, trucks with two trailers, and motorcycles} & \pbox{6cm}{Specific domain knowledge (vehicle shape information) extracted from a laser scanner profile for classification} \\ \hline

    \end{tabular}
\end{table*}

The over-roadway-based vehicle classification systems install sensors over the roadway, offering non-intrusive solutions that do not require physical changes in the roadway, greatly reducing the cost for construction and maintenance. Furthermore, these classification systems are capable of covering multiple lanes and in some cases an entire road segment (\emph{e.g.,} aerial platforms~\cite{cao2011linear}). Since cameras are most widely used sensors for these vehicle classification systems~\cite{chen2012vehicle}\cite{bautista2016convolutional}, the majority part of this section is dedicated to describing the camera-based vehicle classification systems. In this section, we also discuss camera-based classification systems using aerial platforms such as unmanned aerial vehicles (UAVs) and satellites. Although the vehicle classification systems based on cameras have numerous advantages such as the high classification accuracy and the capability of covering multiple lanes, a major downside is the privacy concerns. As such, we discuss a number of privacy-preserving solutions such as the ones based on infrared sensors~\cite{odat2017vehicle}, and laser scanners~\cite{sandhawalia2013vehicle}. Table~\ref{table:over_roadway} summarizes the characteristics of the over-roadway-based vehicle classification systems discussed in this section.

\subsection{Cameras}
\label{sec:over_cameras}

\Figure[t](topskip=0pt, botskip=0pt, midskip=0pt)[width=.95\columnwidth]{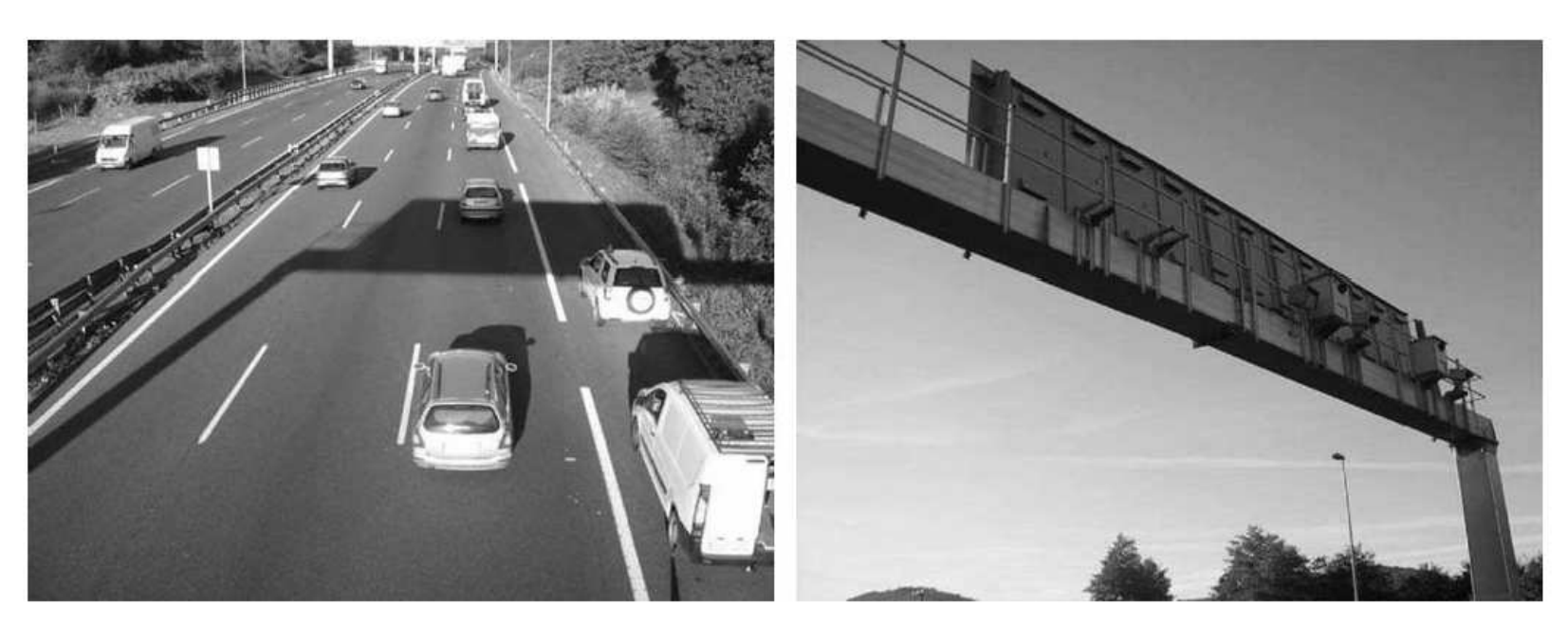}
{A camera-based traffic monitoring system~\cite{unzueta2012adaptive}.\label{fig:camera}}

A most widely adopted sensor for over-roadway-based vehicle classification systems is a camera~\cite{chen2012vehicle}\cite{bautista2016convolutional}. A camera provides rich information for vehicle classification such as the visual features and geometry of passing vehicles ~\cite{tseng2002real}. In comparison with in-road-based vehicle classification systems where multiple sensors are required to cover multiple lanes (\emph{i.e.,} at least a sensor for each lane), a single camera is sufficient for classifying vehicles in multiple lanes (Fig.~\ref{fig:camera}). Advanced image processing technologies supported by sufficient processing power allow for classifying multiple vehicles very quickly and accurately.

The general working of a camera-based vehicle classification system is to capture an image of a passing car, extract features from the image, and run an algorithm to perform vehicle classification. As such, the camera-based systems can be categorized based on how the vehicle image is captured (\emph{e.g.,} methods for reducing the impact of the background image), types of features extracted from the vehicle image, and the mechanisms for performing classification based on the extracted features. A recent trend is that more and more machine learning techniques are applied to extracting features automatically and effectively, and processing the features to build classification models. While earlier systems use simple classification models based on SVM, \emph{k}NN, and decision tree, more advanced machine learning algorithms such as the deep learning are increasingly adopted.

Chen \emph{et al.} focus on effectively capturing a car image from video footage~\cite{chen2012vehicle}. The authors adopt the background Gaussian Mixture Model (GMM)~\cite{zivkovic2006efficient} and the shadow removal algorithm~\cite{chen2009background} to reduce the negative impacts on vehicle classification caused by shadow, camera vibration, illumination changes, \emph{etc}. The Kalman filter is used for vehicle tracking and SVM is used to perform vehicle classification. Experiments were performed with real video footage obtained from cameras deployed in Kingston upon Thames, UK. Vehicles were classified into five categories, \emph{i.e.,} motorcycles, cars, vans, buses, and unknown vehicles. The classification accuracy for these vehicle types was 94.6\%.

Unzueta \emph{et al.} also focus on effectively capturing the car image~\cite{unzueta2012adaptive}. Specifically, the authors address the problem of dynamic changes of the background in challenging environments such as illumination changes and headlight reflections to improve the classification accuracy. A multicue background subtraction method is developed that the segmentation thresholds are dynamically adjusted to account for dynamic changes of the background, and supplementing with extra features extracted from gradient differences to enhance the segmentation~\cite{unzueta2012adaptive}. A two-step approach is proposed to derive spatial and temporal features of a vehicle for classification, \emph{i.e.,} by first generating 2-D estimations of a vehicle silhouette, and then augmenting them to 3-D vehicle volumes for more accurate vehicle classification. Three vehicle types are considered for classification, namely two wheels, light vehicles, and heavy vehicles. The classification accuracy was 92.6\%.

\Figure[t](topskip=0pt, botskip=0pt, midskip=0pt)[width=.95\columnwidth]{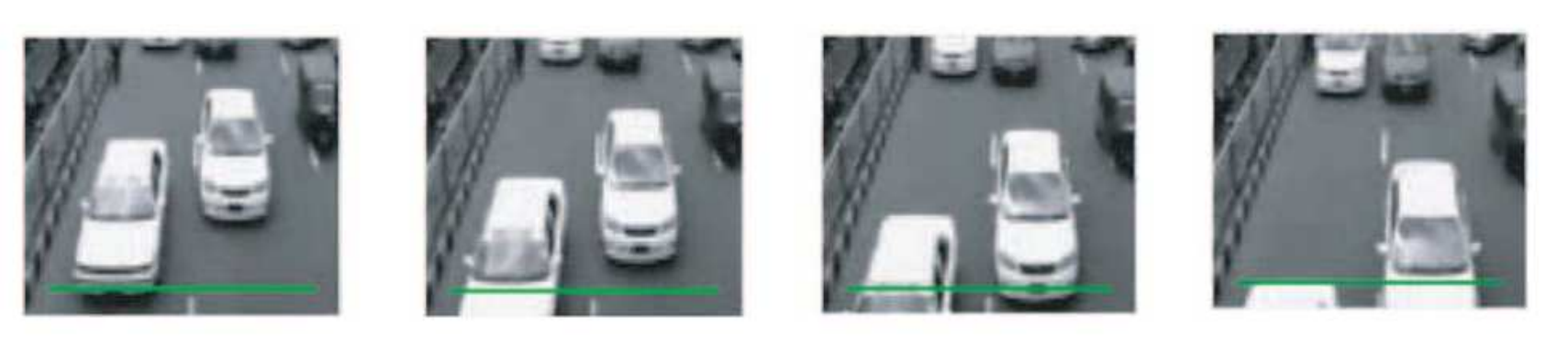}
{An example of the virtual detection lines (VDL)~\cite{mithun2012detection}.\label{fig:vdl}}

Mithun \emph{et al.} propose a multiple virtual detection lines (MVDLs)-based vehicle classification system~\cite{mithun2012detection}. The VDL is a set of line indices of a frame for which the position is perpendicular to the moving direction of a vehicle (Fig.~\ref{fig:vdl}). The pixel strips on a VDL in chronological frames create a time spatial image (TSI). Multiple TSIs are used for vehicle detection and classification to reduce misdetection mostly due to occlusion. Specifically, a two-step process is proposed for classification. Vehicles are first classified into four general types based on the shape-based features. After that, another classification scheme based on the texture-based and shape-invariant features is applied to classify a vehicle into more specific types including motorbikes, rickshaws, autorickshaws, cars, jeeps, covered vans, and busses. The classification accuracy was between 88\% and 91\%

Identifying effective features from the car images is another important challenge for camera-based vehicle classification systems. Karaimer \emph{et al.} combine the shape-based classification and the Histogram of Oriented Gradient (HOG) feature-based classification methods in order to improve the classification performance~\cite{karaimer2015combining}. Specifically, \emph{k}NN is used for the shape-based features including convexity, rectangularity, and elongation, and SVM is used with the HOG features. The two methods are combined using different combination schemes, \emph{i.e.,} the sum rules and the product rules. The sum rule determines the vehicle class such that the sum of the two probabilities for the two classifiers is maximized, and the product rule determines based on the product of the two probabilities. Three vehicle classes were used for experiments, namely, cars, vans, and motorcycles. The classification accuracy was 96.5\%.

Machine learning algorithms are used to extract effective features automatically. Huttunen \emph{et al.} design a deep neural network (DNN) that extracts features from a car image with background, removing the preprocessing steps of detecting a car from an image and aligning a bounding box around the car~\cite{huttunen2016car}. The hyper-parameters of the neural network are selected based on a random search that finds a good combination of the parameters~\cite{bergstra2012random}. The proposed system was evaluated with a database consisting of 6,555 images with four different vehicle types, \emph{i.e.,} small cars, busses, trucks, and vans. The classification accuracy was 97\%.

Dong~\emph{et al.} applies the semisupervised convolutional neural network (CNN) for feature extraction~\cite{dong2015vehicle}. In this work, vehicle front view images are used for classification. Specifically, the proposed CNN consists of two stages. In the first stage, the authors design an unsupervised learning mechanism to obtain the effective filter bank of CNN to capture discriminative features of vehicles. In the second stage, the Softmax classifier is trained based on the multi-task learning~\cite{kumar2012learning} to provide the probability for each vehicle type. Experiments were conducted with two data sets, \emph{i.e.,} the BIT-Vehicle data set~\cite{bit_dataset}, and the data set used by Peng \emph{et al.}~\cite{peng2012vehicle}. The former data set consists of 9,850 vehicle images with six types: bus, microbus, minivan, sedan, SUV, and truck; the latter includes 3,618 daylight and 1,306 nighttime images with truck, minivan, bus, passenger car, and sedan. The classification accuracy for the two data sets were 88.1\%, and 89.4\%, respectively.

In line of the research based on advanced machine learning techniques, Adu-Gyamfi \emph{et al.} develop a vehicle classification system using the deep convolutional neural network (DCNN) that is designed to extract vehicular features quickly and accurately~\cite{adu2017automated}. Compared to other approaches, the DCNN model is pretrained with an auxiliary data set~\cite{russakovsky2015imagenet} and then is fine-tuned with the domain specific data collected from the Virginia and Iowa DOT CCTV camera database. The vehicles were classified into FHWA's 13 vehicle types. The results show that the classification accuracy was greater than 89\%.

Although machine learning techniques make the feature extraction process more effective and consequently improve the vehicle classification accuracy, numerous challenges still remain to be addressed. One of those challenges is to classify visually similar vehicles. Javadi \emph{et al.} propose to apply the fuzzy c-means (FCM) clustering~\cite{bezdek2013pattern} based on vehicle speed as an additional feature to address this challenge~\cite{javadi2018vehicle}. Specifically, they exploit the prior knowledge about varying traffic regulations and vehicle speeds to enhance the classification accuracy for the vehicles with similar dimensions. The proposed classification approach was evaluated with the vehicle images collected for 10 hours from a real highway, classifying the vehicles into four types, namely private cars, light trailers, buses, and heavy trailers. The classification accuracy of 96.5\% was achieved.

Another challenge for applying machine learning techniques for automating background processing and feature extraction is that different parts of an image of a passing car are treated without distinctions, degrading the performance~\cite{krizhevsky2012imagenet}\cite{sivaraman2013looking}. Zhao \emph{et al.} focus on this problem that potentially misses the key part of a car image~\cite{zhao2017deep}. Their work is motivated by the human vision system that distinguishes the key parts of an image from the background, which is called the multiglimpse and visual attention mechanism~\cite{rensink2000dynamic}. This remarkable capability of focusing on only the relevant part of the image allows the human to classify images very accurately. The key idea of their work is thus to exploit the visual attention mechanism to generate a focused image first and provide the image as input to CNN for more accurate vehicle classification. They performed experiments to classify a vehicle into five types, sedans, vans, trucks, SUVs, and coaches, and achieved the classification accuracy of 97.9\%.

Theagarajan \emph{et al.} observe that machine learning algorithms work only effectively with an extremely large amount of image data~\cite{theagarajan2017eden}. The authors also find that most camera-based classification systems are built upon small traffic data sets that do not take into account sufficiently the variability in weather conditions, camera perspectives, and roadway configurations. To address this problem, they develop a deep network-based vehicle classification mechanism utilizing the largest data set that is ever known to the research community. The data set contains 786,702 vehicle images from cameras at 8,000 different locations in USA and Canada. With the huge amount of data, they classified vehicles into 11 types including articulated trucks, background, busses, bicycles, cars, motorcycles, nonmotorized vehicles, pedestrians, pickup trucks, single unit trucks, and work vans. They obtained high classification accuracy of 97.8\%.

The same dataset~\cite{theagarajan2017eden} was used by Kim and Lim~\cite{kim2017vehicle}. Different from other works based on CNN, the authors apply a data augmentation technique to enhance the performance under different sample sizes for different types of cars. The authors also apply a weighing mechanism that associates a weight depending on different vehicle types. The classification accuracy was 97.8\%. The imbalanced dataset problem was also addressed by Liu \emph{et al.}~\cite{liu2017ensemble}. Specifically, to increase the number of samples for certain vehicle types, they apply various data augmentation techniques such as random rotation, cropping, flips, and shifts and created an ensemble of CNN models based on the parameters obtained from the augmented dataset. The proposed work was tested with the MIO-TCD classification challenge dataset which classifies the vehicles into 11 types. They achieved the classification accuracy of 97.7\%.

The vehicle occlusion problem is another challenge for applying machine learning algorithms to camera-based vehicle classification. Chang \emph{et al.} propose an effective model based on the Recursive Segmentation and Convex Hull (RSCH) to address this problem~\cite{chang2018vision}. Specifically, vehicles are assumed as convex regions, and a decomposition optimization model is derived in order to separate vehicles from a multi-vehicle occlusion. After addressing the occlusion problem, vehicle classification is performed with a regular CNN. Experiments were conducted with the CompCars dataset~\cite{yang2015large} which consists of 136,726 vehicle images with five types: sedans, SUVs, vans, busses, and trucks. For this dataset, the authors achieved the classification accuracy of 97.6\%.

Some vehicle classification systems integrate a camera with a different type of sensor. Hasnat \emph{et al.} significantly improve the classification accuracy by integrating a camera with optical sensors~\cite{hasnat2018new}. They call it a hybrid classifier system. Specifically, the system consists of both the optical sensor-based classifier and the CNN-based classifier. And then, they apply the Gradient Boosting technique~\cite{friedman2001greedy} to combine the decisions from these classifiers, constructing a stronger predictor based on the base predictors. Five vehicle classes are defined for classification: light vehicles (height less than 2m), intermediate vehicles (height between 2m and 3m), heavy vehicles (height greater than 3m), heavy vehicles with more than 2 axles, and motorbikes. The classification accuracy was 99.0\%.

\subsection{Aerial Platforms}
\label{sec:aerial}

\Figure[t](topskip=0pt, botskip=0pt, midskip=0pt)[width=.95\columnwidth]{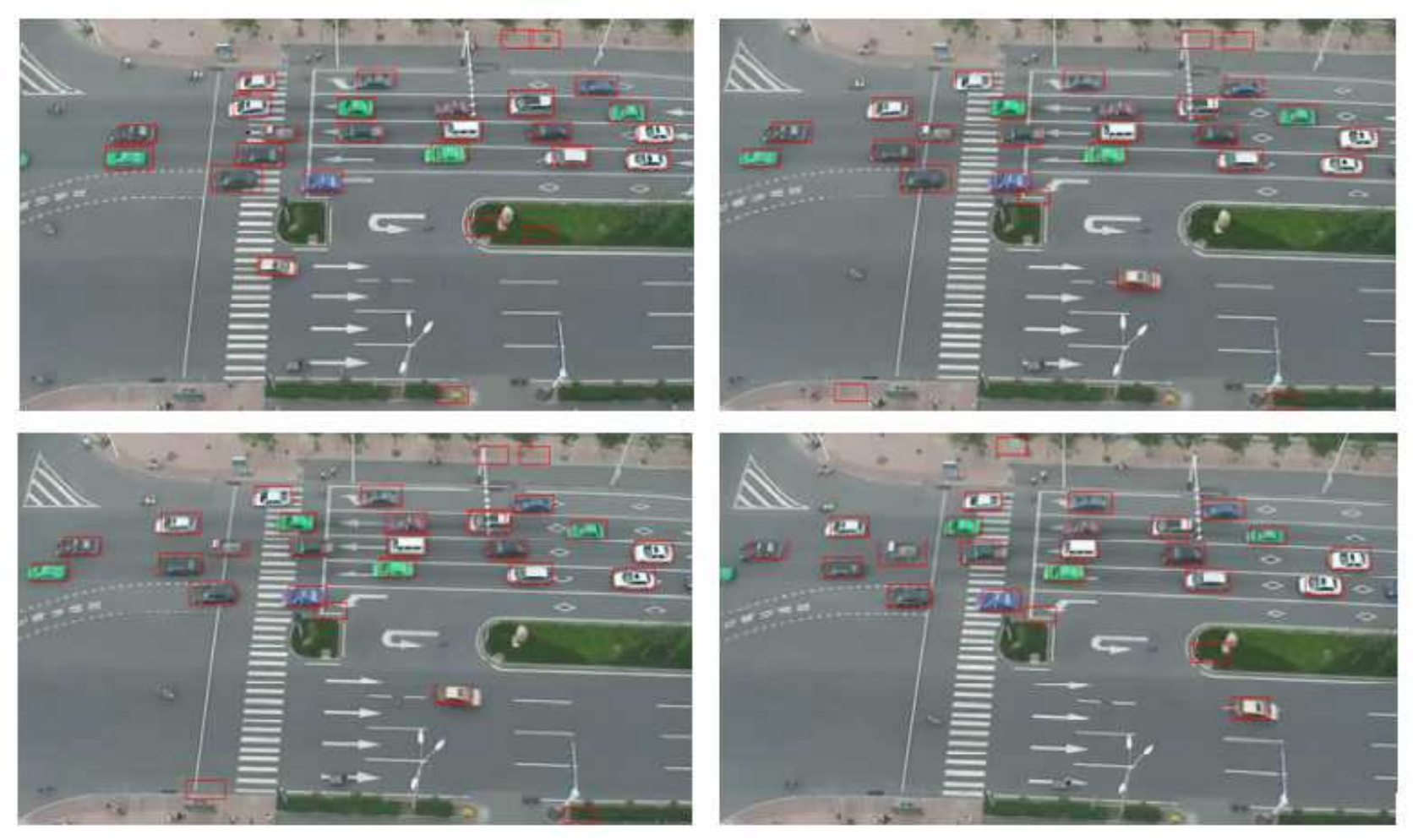}
{An example of an aerial image and vehicle detection using SVM~\cite{cao2011linear}.\label{fig:aerial_image}}

Cameras are mounted on aerial platforms such as UAVs and satellites in order to cover wide areas (e.g., an entire roadway segment) for vehicle classification (Fig.~\ref{fig:aerial_image}). Despite the advantage of wide coverage, vehicle classification based on aerial platforms is a non-trivial task due to the low image resolution. In fact, even detecting a vehicle is not an easy task for aerial platform-based systems. For example, Cao \emph{et al.} develop a method for vehicle detection based on an airborne platform~\cite{cao2011linear}. Their main idea to enhance the vehicle detection accuracy is to exploit a new feature called the boosting HOG and perform classification using the linear SVM. Videos were captured in an urban traffic environment to evaluate the proposed system. While most ground-based traffic monitoring systems achieve near 99\% accuracy for vehicle detection (note that this is not for vehicle classification), their system achieved the vehicle detection accuracy of 90\%.

Due to the low image resolution, many aerial platform-based vehicle classification systems aim to classify only for a limited number of vehicle types that are relatively easily distinguishable such as cars and trucks. Liu and Mattyus develop a vehicle classification system based on an aerial platform especially focusing on reducing the computation speed for vehicle classification~\cite{liu2015fast}. A binary sliding window detector is applied to detect a vehicle from an aerial image. Once a vehicle is detected, the HOG features are extracted from the image of the detected vehicle~\cite{dalal2005histograms} using a neural network with a single hidden layer~\cite{lecun2012efficient}. Vehicles are then classified into two types, \emph{i.e.,} cars and trucks. The classification accuracy was 98.2\% which is relatively high due to the small number of easily distinguishable vehicle types for classification.

With the help of advanced machine learning techniques, the classification accuracy of some aerial platform-based vehicle classification systems is improved. However, the results are not yet comparable to ground sensor-based vehicle classification systems. Tan \emph{et al.} develop a two-step vehicle classification method using aerial images~\cite{tan2018vehicle}. A change detection scheme is applied to detect vehicles based on pixel-level changes represented as a heat map. And then, a standard CNN is used for classification. In particular, they adopt the fully connected layer of the AlexNet model~\cite{krizhevsky2012imagenet}, and the final classification layer of the Inception model~\cite{szegedy2016rethinking}. Experiments were performed with the images collected from a manned aircraft. The vehicles were classified into four classes: sedans, vans, pickups, and trucks. The classification average accuracy was 80.3\%.

Audebert \emph{et al.} also apply a standard CNN to aerial images for vehicle classification~\cite{audebert2017segment}. Various CNN models are adopted such as LeNet~\cite{lecun1998gradient}, AlexNet~\cite{krizhevsky2012imagenet}, and VGG-16~\cite{simonyan2014very} pre-trained with existing training datasets. To overcome the discrepancy between the training datasets and testing datasets, the authors utilize data normalization and augmentation techniques based on the geometric operations including translations, zooms, rotations of images. The experiments were performed with the NZAM/ONERA Christchurch dataset classifying the vehicles into cars, vans, pickups, trucks. The highest classification accuracy of 80\% was achieved with the VGG-16 model.

\subsection{Privacy Preserving Solutions}
\label{sec:infrared}

\Figure[t](topskip=0pt, botskip=0pt, midskip=0pt)[width=.8\textwidth]{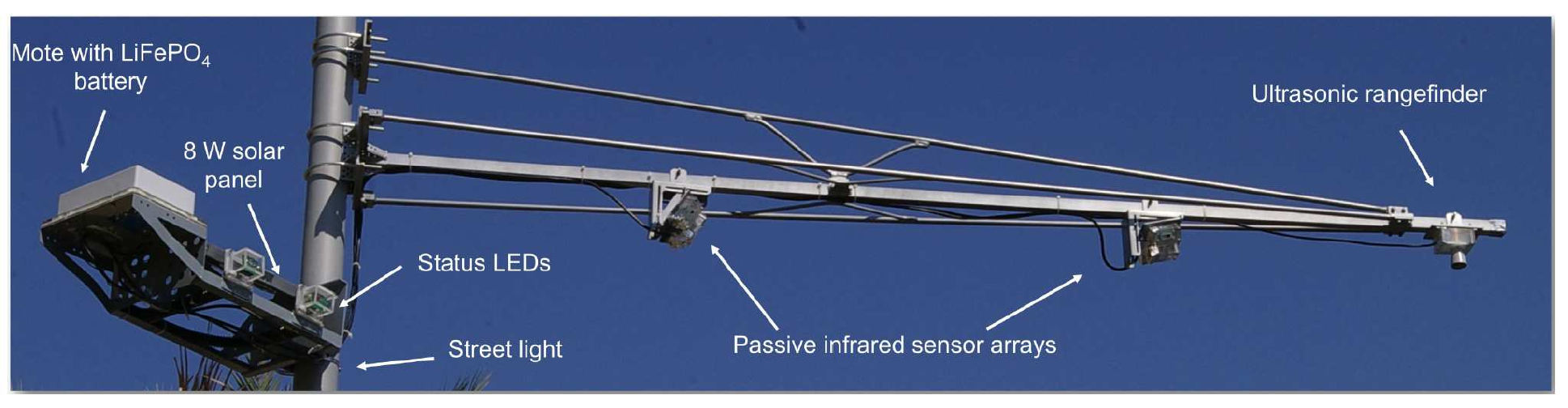}
{Vehicle classification based on infrared and ultrasonic sensors~\cite{odat2017vehicle}.\label{fig:infrared}}

A major downside of camera-based vehicle classification systems including aerial platform-based systems is the privacy concerns. Various privacy preserving solutions have been developed using different kinds of sensors. Odat \emph{et al.} propose a system based on the combination of infrared and ultrasonic sensors~\cite{odat2017vehicle} (Fig.~\ref{fig:infrared}) The Bayesian network and neural network are used to fuse extracted features from sensor data collected from both sensors. Specifically, the heights of different parts of a passing vehicle measured based on the ultrasonic sensor are used as key features. Also, other features extracted from the infrared sensors, \emph{i.e.,} the inverse of the estimated delay and the estimated duration are used for classification.  Vehicles were classified into sedan, pickup truck, SUV, bus, and two wheeler. The highest classification accuracy was 99\%.

Sandhawalia \emph{et al.} develop a privacy preserving solution using the laser scanners~\cite{sandhawalia2013vehicle}. The laser scanners perform 3D scan of the vehicle surface allowing for accurate estimation of the width, height, and length of a passing vehicle. It is noted that although the laser scanners address the privacy concerns, they are sensitive to extreme weather conditions and the cost for installation is higher than cameras. The authors represent a laser scanner profile as an image to perform image classification. Specifically, an image presentation technique, \emph{i.e.,} the Fisher vector~\cite{perronnin2010improving} is applied to extract effective features from a laser scanner image. In this work, vehicles were classified into six types: passenger vehicles, passenger vehicles with one trailer, trucks, trucks with one trailer, trucks with two trailers, and motorcycles. The classification accuracy was 82.5\%.

Another laser scanner-based approach is developed by Chidlovskii \emph{et al.}~\cite{chidlovskii2014vehicle}. The key contribution of this vehicle classification system in comparison with~\cite{sandhawalia2013vehicle} is to utilize the specific domain knowledge, \emph{i.e.,} the vehicle shapes to enhance the classification accuracy. Specifically, vehicle shapes are extracted from the laser scans to analyze a vehicle as a multi-dimensional object. To address the space shift and scaling problem, the dynamic time warping (DTW)~\cite{berndt1994using} and the global alignment kernel (GA)~\cite{cuturi2007kernel} are used. The same six vehicle types as~\cite{sandhawalia2013vehicle} were used for experiments. The best classification accuracy achieved was 86.8\%.

\section{Side-Roadway-Based Vehicle Classification}
\label{sec:side_roadway}

Side-roadway-based vehicle classification systems deploy sensors on a roadside. Similar to over-roadway-based classification systems, a key advantage of side-roadway-based systems is the capability of covering multiple lanes simultaneously. Additionally, side-roadway-based systems are easier to install quickly at a reduced cost as no traffic disturbance and lane closure is needed at all, which makes these systems especially appropriate for ad-hoc monitoring purposes. However, a critical challenge lies in detecting and classifying overlapping vehicles because it is extremely challenging to collect sensor data corresponding to occluded vehicles, and even the sensor data collected for front vehicles are easily distorted due to the occluded vehicles. Various kinds of sensors are used to implement side-roadway-based systems such as the magnetic sensors~\cite{taghvaeeyan2014portable}, acoustic sensors~\cite{ntalampiras2018moving}, LIDAR~\cite{asborno2019truck}, radar~\cite{raja2016analysis}, radio tranceivers~\cite{sliwa2018leveraging}, and Wi-Fi transceivers~\cite{won2017witraffic}. Table~\ref{table:sideroadway} summarizes the characteristics of these side-road-based vehicle classification systems.

\begin{table*}[htbp]
    \caption{Side-roadway-based vehicle classification systems}
    \label{table:sideroadway}
    \center
     \begin{tabular}{| l | l | l | l | l |}
    \hline
    Major Equipment & Publications & Accuracy & Vehicle Classes & Key Features\\ \hline

    \pbox{3cm}{Magnetic sensors} & \pbox{3cm}{Taghvaeeyan TITS'14~\cite{taghvaeeyan2014portable}} & \pbox{1cm}{83.0\%} & \pbox{3cm}{Class I (sedan), class II (SUV, pickup, van), class III (bus, two-three-axle trucks), class IV (articulated bus, four-to-six-axle truck)} & \pbox{6cm}{Magnetic height as a key feature to address the problem of classifying vehicles with the same length} \\ \cline{2-5}

    \pbox{3cm}{} & \pbox{3cm}{Wang TITS'14~\cite{wang2014easisee}} & \pbox{1cm}{93.0\%} & \pbox{3cm}{bicycles (including bicycles, electric bicycles and motorcycles), cars (including family cars, taxis, and SUVs), and minibuses} & \pbox{6cm}{Magnetic sensor used for collaborative sensing with a camera to reduce power consumption} \\ \cline{2-5}

    \pbox{3cm}{} & \pbox{3cm}{Yang IEEE Sensors'15~\cite{yang2015vehicle}} & \pbox{1cm}{93.6\%} & \pbox{3cm}{motorcycle, two-box, saloon, bus and sport utility vehicle (SUV)} & \pbox{6cm}{Vehicle classification for low-speed congested traffic} \\ \hline


   \pbox{3cm}{Acoustic sensors} & \pbox{3cm}{Bischof IS'10~\cite{bischof2010autonomous}} & \pbox{1cm}{85.0\%} & \pbox{3cm}{cars and trucks} & \pbox{6cm}{Acoustic sensors used to support autonomous training for the camera-based system} \\ \cline{2-5}

   \pbox{3cm}{} & \pbox{3cm}{Ntalampiras TETCI'18~\cite{ntalampiras2018moving}} & \pbox{1cm}{96.3\%} & \pbox{3cm}{assault amphibian vehicle
(AAV) and dragon wagon (DW)} & \pbox{6cm}{A group of acoustic sensors; Sensor-specific classification model; faulty sensor detection} \\ \hline

    \pbox{3cm}{Lidar} & \pbox{3cm}{Lee TRR'12~\cite{lee2012side}, JITS'15~\cite{lee2015using}} & \pbox{1cm}{99.5\%} & \pbox{3cm}{motorcycle, passenger vehicle, passenger vehicle pulling a trailer, single-unit truck, single-unit truck pulling a trailer, and multiunit truck} & \pbox{6cm}{Vehicle body information (vehicle length and height) extracted from accurate LiDAR data used as key features} \\ \cline{2-5}

   \pbox{3cm}{} & \pbox{3cm}{Asborno TRR'19~\cite{asborno2019truck}} & \pbox{1cm}{96.0\%} & \pbox{3cm}{ van and container, platform, low-profile trailer, tank, and hopper and end dump} & \pbox{6cm}{Designed specifically for classification of truck body types; The duration and vehicle body points are the main features} \\ \hline

   \pbox{3cm}{RF Transceivers} & \pbox{3cm}{Haferkamp VTC'17~\cite{haferkamp2017radio}} & \pbox{1cm}{99.0\%} & \pbox{3cm}{passenger cars, trucks} & \pbox{6cm}{Received signal strength (RSSI) as a key feature; kNN and SVM for classification} \\ \cline{2-5}

   \pbox{3cm}{} & \pbox{3cm}{Silwa ITSC'18~\cite{sliwa2018leveraging}} & \pbox{1cm}{89.1\%} & \pbox{3cm}{passenger cars, passenger
cars with trailer, SUVs, minivans, vans, trucks, truck with
trailers, buses, and transporters} & \pbox{6cm}{Multiple sets of RF transmitters and receivers} \\ \hline

   \pbox{3cm}{Radar} & \pbox{3cm}{Raja Sensors'16~\cite{raja2016analysis}} & \pbox{1cm}{99.0\%} & \pbox{3cm}{compact, saloon and small sport utility vehicle (SUV)} & \pbox{6cm}{Power spectral density of the time-domain signal as input to kNN; The classification accuracy depends on the distance between the radio receiver and the passing car} \\ \hline


   \pbox{3cm}{Wi-Fi Transceivers} & \pbox{3cm}{Won ICCCN'17~\cite{won2017witraffic}} & \pbox{1cm}{96.0\%} & \pbox{3cm}{passenger vehicles, and trucks} & \pbox{6cm}{The first Wi-Fi-based traffic monitoring system that is build upon a pair of Wi-Fi transceivers to reduce the cost.} \\ \cline{2-5}

   \pbox{3cm}{} & \pbox{3cm}{Won ArXiv'18~\cite{won2018deepwitraffic}} & \pbox{1cm}{91.1\%} & \pbox{3cm}{motorcycle, passenger car, SUV, pickup truck, large truck} & \pbox{6cm}{A Wi-Fi-based traffic monitoring system with an advanced machine learning technique to enable classification for more vehicle types.} \\ \hline

    \end{tabular}
\end{table*}

\subsection{Magnetic Sensors}
\label{sec:side_magnetic}

As we discussed in Section~\ref{sec:in_roadway}, magnetic sensors have been widely adopted in in-road-based vehicle classification systems. These magnetic sensors are also frequently adopted in side-road-based vehicle classification systems especially to overcome one of the critical limitations for in-road-based systems, that is the high cost for installation and maintenance. While the basic mechanism for these side-roadway-based classification systems is similar to the in-road-based systems in that vehicle classification is performed based on the magnetic profile of a passing car, numerous research challenges are addressed such as classifying vehicles with similar body sizes (\emph{e.g.,} SUVs and pickup trucks), and classifying overlapping vehicles effectively.

Taghvaeeyan \emph{et al.} develop a vehicle classification system based on three-axis magnetic sensors (Fig.~\ref{fig:amr}) deployed roadside focusing on addressing the problem of classifying vehicles with  similar body sizes~\cite{taghvaeeyan2014portable}. The key idea is to utilize both the vehicle length and height as the main features for vehicle classification. More precisely, while existing in-road-based systems based on magnetic sensors measure only the vehicle length, their system is capable of obtaining the vehicle height information by placing another magnetic sensor above a magnetic sensor roadside and measuring the ratio of the sensor readings from the two sensors. Vehicles were classified into five categories: Class I (sedans), Class II (SUVs, pickups, and vans), Class III (buses, two- and three-axle trucks). Class IV (articulated buses and four- to six-axle trucks). The classification accuracy was 83\%.

\Figure[t](topskip=0pt, botskip=0pt, midskip=0pt)[width=.75\columnwidth]{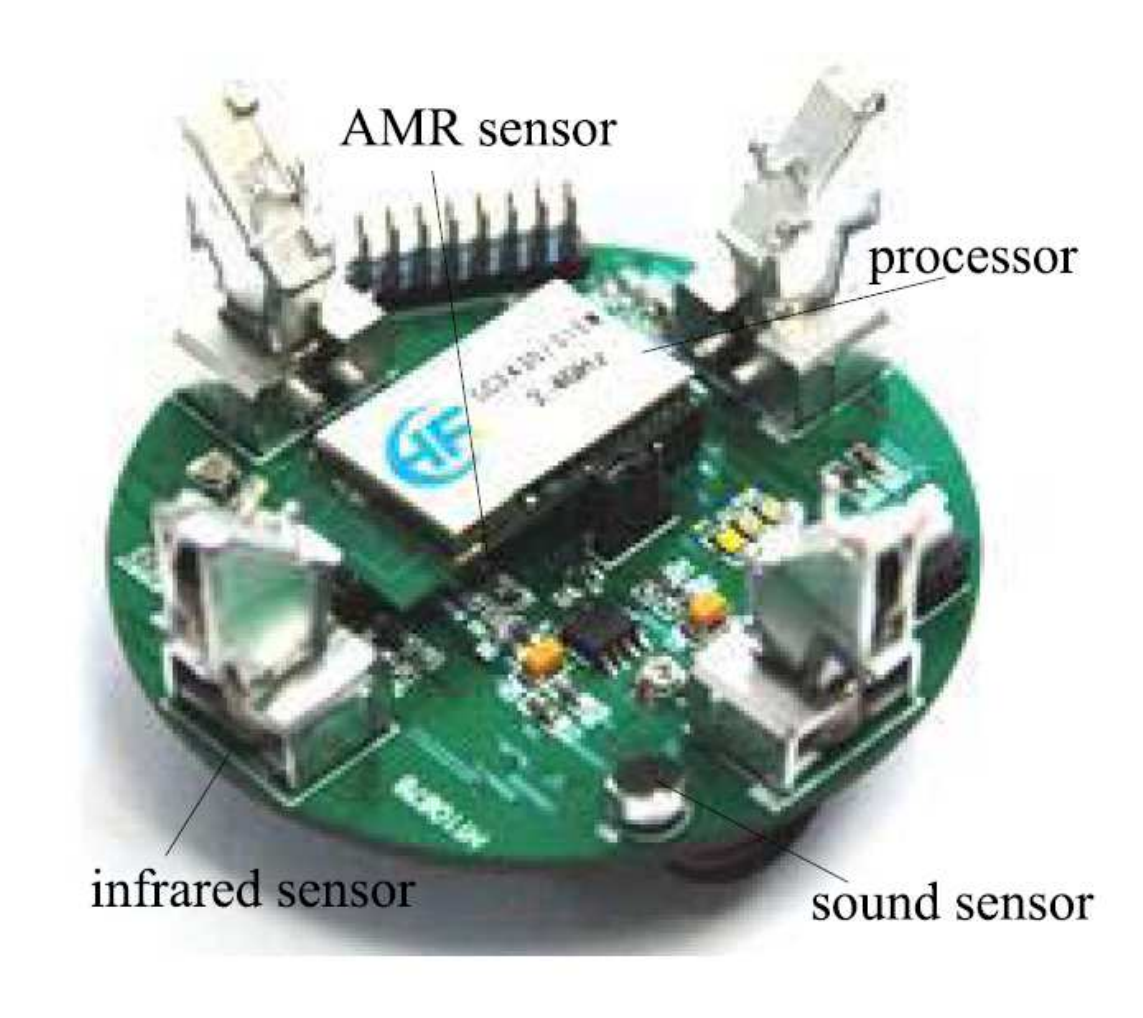}
{The three-axis AMR sensor used by~\cite{taghvaeeyan2014portable}.\label{fig:amr}}

Yang and Lei focus on another interesting problem for magnetic sensor-based vehicle classification systems, \emph{i.e.,} classifying vehicles that are very close to each other, which typically happens under low-speed congested traffic conditions~\cite{yang2015vehicle}. When vehicles are too close, the magnetic signals are significantly distorted making the vehicle classification process extremely challenging. To address this problem, the authors propose a hierarchical tree-based approach~\cite{kaewkamnerd2009automatic}. The key idea is to identify and extract effective features from the magnetic signal that are immune to signal distortions caused by the small inter-vehicle distance. Specifically, five features including the signal duration, signal energy, average energy, and ratios of the positive and negative energy are extracted. A hierarchical tree is constructed by comparing the values of these features, which is then used to classify vehicles into five categories: motorcycle, two-box, saloon, bus and sport utility vehicle (SUV). The classification accuracy was 93.6\%.

In some cases, magnetic sensors are used for collaborative sensing. EasiSee is basically a camera-based vehicle classification system~\cite{wang2014easisee}. A key motivation for integrating a magnetic sensor with the system is to reduce the power consumption of the system. Specifically, the magnetic sensor is used to detect a passing vehicle, and only when a vehicle is detected, the camera is activated, thereby being able to put the camera in the sleep mode, reducing power consumption. They also develop an efficient image processing algorithm focusing on reducing the computational complexity. Vehicles were classified into bicycles (including bicycles, electric bicycles and
motorcycles), cars (including family cars, taxis, and SUVs), and minibuses. The classification accuracy was 93\%.


\subsection{Acoustic Sensors}
\label{sec:side_acoustic}


The acoustic sensor-based vehicle classification systems capture the audio signal induced by a passing vehicle using microphone sensors. The success of these types of solutions depends largely on effective feature extraction from acoustic signals. However, since the performance of the acoustic sensors are easily affected by ambient noise, it is very challenging to identify such effective features. As a result, the acoustic sensors are typically used to support operation of other types of sensors such as cameras~\cite{bischof2010autonomous}. Additionally, a group of acoustic sensors are deployed to mitigate the impact of ambient noise and increase the classification accuracy~\cite{ntalampiras2018moving}.

Bischof \emph{et al.} adopt an acoustic sensor to support the self learning process of a camera-based vehicle classification system~\cite{bischof2010autonomous}. The proposed system consists of audio-based and video-based classification systems. The audio-based system acts as a supervisor to enable autonomous training of the video-based system, obviating the needs for manually labeling the huge amount of video data. Specifically, the audio sensor-based system performs a priori classification for a passing car and forwards the classification results with a confidence level to the video-based system. And then, the video-based system uses the results for autonomously training the classification model. The proposed system was evaluated with different kinds of classifiers such as \emph{k}NN, SVM, and ANN.  Vehicles were classified into two types trucks and cars. The classification accuracy was 85\% for trucks and 71\% for cars.

Ntalampiras~\cite{ntalampiras2018moving} develop a wireless acoustic sensor network (WASN) that consists of multiple wireless microphone nodes to make the system resilient to environmental noise. An interesting aspect of their work is that the sensor specific classification models are created at the sensor level, and then the decisions are combined at the higher level using the correlation-based dependence graph. In addition, a stationary checking algorithm is proposed to detect sensor faults, taking advantage of multiple acoustic sensors. Experiments were conducted with the DARPA/IXOs SensIT dataset which consists of two vehicle types, Assault Amphibian Vehicle
(AAV) and Dragon Wagon (DW)~\cite{duarte2004vehicle}. The average classification accuracy was 96.3\%.

\subsection{LIDAR}
\label{sec:side_lidar}

\Figure[t](topskip=0pt, botskip=0pt, midskip=0pt)[width=.95\columnwidth]{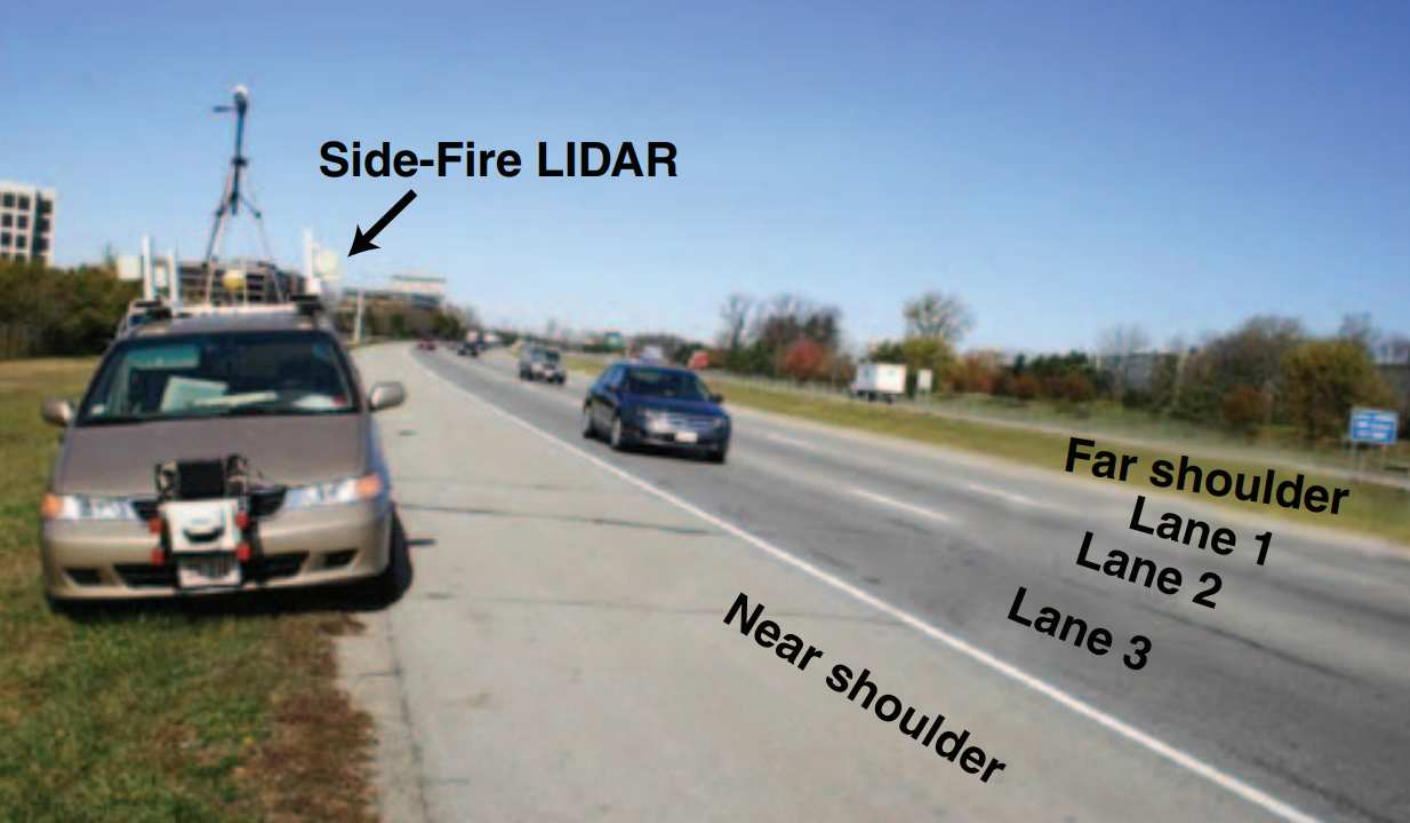}
{An example of a LIDAR-based vehicle classification system~\cite{lee2012side}.\label{fig:lidar_example}}

A light detection and ranging (LIDAR) sensor sends eye-safe laser lights and record the reflections to calculate the points of the environment such as the road, passing vehicles, and vegetation, \emph{etc.} Based on the collected data, effective features are extracted such as the size and shape of a passing car to perform vehicle classification. LIDAR is especially powerful in identifying the shape of a passing car due to its high precision sensing. However, the vehicle occlusion problem remains as a critical challenge for LIDAR-based vehicle classification systems.

Lee and Coifman develop a LIDAR-based vehicle classification system~\cite{lee2012side}\cite{lee2015using}. Two LIDAR sensors that are mounted on the driver side of a car are deployed roadside to scan the body of a passing car vertically (Fig.~\ref{fig:lidar_example}). Specifically, six features are identified and extracted from the LIDAR data which include the vehicle height, vehicle length, middle drop, height at middle drop, front vehicle height, front vehicle length, rear vehicle height, and rear vehicle length. Especially, the middle drop feature is used to classify vehicles pulling trailers; The different height at the middle drop is used to differentiate between passenger vehicles with trailers and trucks with trailers. A classification tree is built by comparing the values of those features. Six vehicle classes were used for classification, \emph{i.e.,} the motorcycle, passenger vehicle, passenger vehicle pulling a trailer, single-unit truck or bus, single-unit truck or bus pulling a trailer, and multi-unit truck. They achieved the classification accuracy of 99.5\%.

Asborno \emph{et al.} focus on classification of truck body types~\cite{asborno2019truck}. Two LIDAR units are deployed roadside. Two key features are defined, \emph{i.e.,} the duration and array of the vehicle body points. The duration means the elapsed time while a passing vehicle was in front of the LIDAR unit, and the vehicle body points capture the shape of the truck body. Based on these two key features as input to several classifiers such as Decision Tree (DT), artificial neural network (ANN), support vector machine (SVM), and Naive Bayes (NB), vehicle classification was performed. The proposed system was deployed at an interstate location to classify vehicles into five different truck body types, \emph{i.e.,} five-axle tractor-trailers (van and container, platform, low-profile trailer, tank, and hopper and end dump). They obtained the classification accuracy up to 96\%.

\subsection{Radar}
\label{sec:side_radar}

The basic mechanism of radar-based vehicle classification systems are similar to LIDAR-based systems in that the radar-based systems exploit the reflections of radio signals from the vehicle body to perform classification. The difference is that while the LIDAR sensors use laser beams, the radar sensors use radio waves. The radar sensors are less vulnerable to weather and light conditions than LIDAR, but the LIDAR sensors provide more accurate representation of the vehicle body.

Raja \emph{et al.} use the passive forward scattering radar (FSR) for vehicle classification~\cite{raja2016analysis}. The radar cross section information is analyzed in the time domain for de-noising and normalization. And then, the power spectral density (PSD) of the time-domain signal is calculated using the Welch algorithm~\cite{welch1967use}. The power spectral density estimates the power of the signal at different frequencies, which is used as input to a classifier. The large data size of the spectral signature of PSD is reduced using the Principle Components Analysis (PCA). After that, \emph{k}NN is applied to classify vehicles into three types: compact, saloon and small sport utility vehicle (SUV). The results indicate that the classification accuracy is significantly influenced by the distance between the receiver and the car, \emph{i.e.,} the classification accuracy was 99\% for 5m, and 82.1\% for 20m.

\subsection{RF Transceivers}
\label{sec:side_radio}

The propagation of radio frequency (RF) signals is disturbed by a passing vehicle. Specifically, a RF transmitter and a receiver are deployed on the opposite sides of a road. When a car passes, the line of sight between the transmitter and the receiver is interrupted resulting in attenuation and reflection of the RF signals. Consequently, distinctive patterns of the received RF signals depending on the shape and size of the passing car are captured by the receiver. These unique patterns are used to classify the vehicles.

Haferkamp \emph{et al.} focus on the attenuation of the RF signal due to a passing car and uses it as a key feature for vehicle classification~\cite{haferkamp2017radio}. The signal attenuation is represented by the received signal strength indicator (RSSI). The RSSI traces corresponding to the passing vehicle are provided as input to classifiers, \emph{i.e.,} \emph{k}NN and SVM. A five-fold cross validation is used to perform classification. The vehicles were classified into passenger cars and trucks. The classification accuracy was 99\% which is quite high due to the small number of vehicle types.

Silwa \emph{et al.} utilize the the low-rate wireless personal area networks (LR-WPANs), \emph{i.e.,} the IEEE 802.15.4 standard to capture the radio fingerprint of a passing vehicle for vehicle classification~\cite{sliwa2018leveraging}. Similar to~\cite{haferkamp2017radio}, RSSI is used as the main feature, while the proposed system is designed to achieve more accurate and reliable vehicle classification. Specifically, three transmitters and three receivers are deployed on each side of the street with the fixed longitudinal distances. Three different classifiers are adopted, \emph{i.e.,} SVM~\cite{cortes1995support}, CNN~\cite{lecun1989generalization}, and Random Forests (RF)~\cite{breiman2001random}. The system classifies vehicles into 9 different types: passenger cars, passenger cars with trailer, SUVs, minivans, vans, trucks, truck with trailers, buses, and transporters. The average classification accuracy was 89.1\%.

\subsection{Wi-Fi Transceivers}
\label{subsec:wifi}

Recently, Wi-Fi-based traffic monitoring systems have been developed specifically targeting the endemic cost issue for deploying a large number of traffic monitoring systems to cover huge miles of rural highways. The idea is to leverage the unique Wi-Fi channel state information (CSI) patterns~\cite{halperin2011tool} induced by passing vehicles to perform vehicle classification. Specifically, the spatial and temporal correlations of CSI phase and amplitude enable effective vehicle classification. Especially the significantly low cost of off-the-shelf Wi-Fi transceivers enable large-scale deployment of traffic monitoring systems. Won \emph{et al.} develop the first prototype system and demonstrate the average vehicle classification accuracy of 96\%~\cite{won2017witraffic}. However, the prototype classifies vehicles only into passenger cars and trucks. The authors, in their extended version of the work, applies an advanced machine learning technique, \emph{i.e.,} a convolutional neural network (CNN) to extract the effective features of the CSI data automatically and enables classification for more vehicle types including motorcycles, passenger cars, SUVs, pickup trucks, and large trucks~\cite{won2018deepwitraffic}. They achieved the average classification accuracy of 91.1\%.

 Specifically, a convolutional neural network
(CNN) is designed to capture the optimal features of CSI
data automatically and train the vehicle classification model
based on effectively preprocessed CSI data as input. Numerous
techniques are applied to address challenges of improving the
classification accuracy.

\section{Challenges for Future Research}
\label{sec:future_research}


We have witnessed significant development of vehicle classification systems in the past decade. Thanks to recent advances in sensing, machine learning, and wireless communication technologies, the classification accuracy has improved greatly at a significantly reduced cost. However, these emerging vehicle classification systems have left a number of open questions. In this section, we discuss these challenges and several future research directions.

\Figure[t](topskip=0pt, botskip=0pt, midskip=0pt)[width=.95\columnwidth]{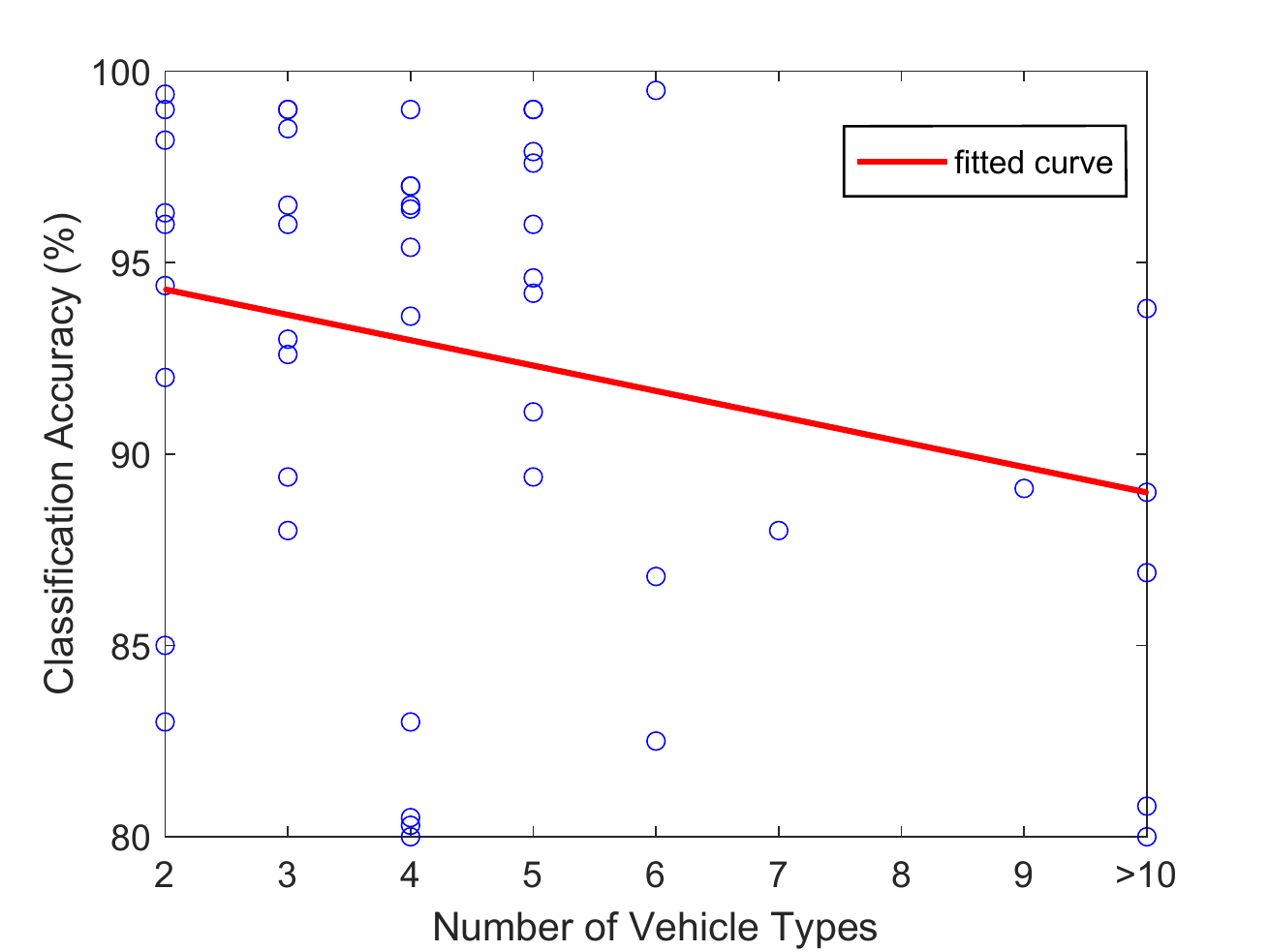}
{The classification accuracy for different numbers of vehicle types.\label{fig:acc_vs_type}}

First of all, we need to define a standard that defines a list of vehicle types for classification to allow various vehicle classification systems to be evaluated based on the same set of vehicle types, so that the system developers and researchers will be able to evaluate the performance of their systems more effectively and fairly. Additionally, the standard will enable the users such as the government agencies to make more informed decision on selecting most appropriate vehicle classification systems. Unfortunately, however, various vehicle classification systems have been evaluated with extremely different vehicle types. Fig.~\ref{fig:acc_vs_type} displays the classification accuracy for different numbers of vehicle types of the vehicle classification systems that we review in this article. The fitted curve in this figure indicates that the vehicle classification systems that are evaluated with a smaller number of vehicle types tend to show higher classification accuracy, although such high classification accuracy is not guaranteed when the system is applied to a larger number of vehicle types.

Another important issue that makes fair comparison of vehicle classification systems difficult is the varying experimental conditions used for evaluating the vehicle classification systems. There are numerous factors that should be controlled to allow for fair comparison of the performance such as the number of lanes, obstacles, and weather conditions. For example, weather conditions affect the performance of certain types of sensors such as the camera, LIDAR, radio, and Wi-Fi. Side-firing sensors are significantly affected by the number of lanes due to overlapped vehicles. Some sensors such as the acoustic sensors are exceptionally vulnerable to ambient noise. To address this challenge, an universally accepted standard for experimental configurations can be developed.

Vehicle classification systems should conform to a common set of performance metrics. However, numerous vehicle classification systems focus only on measuring the classification accuracy while ignoring other performance metrics such as the cost for maintenanace/installation, the capability of classifying overlapped vehicles, sustainability (duration of operation), and resiliency to weather conditions/noise. For example, while camera-based classification systems achieve high classification accuracy, these systems suffer from the privacy concerns. Similarly, many in-road-based classification systems have high classification accuracy due to close contact with passing cars, but these systems are very costly to install and maintain.

One of the critical challenges especially for side-roadway-based classification systems is the vehicle occlusion problem. The operation of numerous kinds of sensors such as the magnetic sensors, LIDAR, Radar, RF, and Wi-Fi is disturbed by occluding vehicles, making it nearly impossible to accurately classify the overlapped vehicles. A possible approach to address this challenge is to take advantages of the over-roadway-based systems to develop a more efficient side-roadway-based systems. Specifically, the side-firing sensors can be placed at different heights so that each sensor can cover each lane explicitly without being interrupted by the vehicles in other lanes. For example, a LIDAR sensor can be configured to record reflections from a targeted lane only. To the best of our knowledge, there is no side-roadway-based vehicle classification systems that consider the better strategy of placing sensors to overcome the vehicle occlusion problem.

More and more vehicle classification systems depend on machine learning techniques. To achieve high classification accuracy, however, a huge amount of data should be collected to train and create an effective classification model. Especially, the manual labeling process for training the classification model requires a significant amount of time and efforts. It also requires extra efforts for obtaining the ground truth data. A possible future research direction is to develop a ``closed loop self-learning'' vehicle classification system. Once deployed, these systems will train the classification models that will autonomously and continuously evolve based on trial and error.

Although we have seen that many classification systems achieve very high classification accuracy, achieving near 100\% classification accuracy especially for a large number of vehicle types is still a very challenging task. One possible reason for the difficulty lies in the fact that most solutions rely on a single type of sensor for vehicle classification. There are few works that utilize the hybrid approach of combining the advantages of different types of sensors, and even the different types of deployment methods, \emph{e.g.,} combination of the side-roadway-based and over-roadway-based systems. These heterogeneous sensor systems will communicate and exchange various kinds of information to offset their weaknesses and capitalize their strengths to achieve higher classification accuracy. For example, the camera-based system may be adaptively controlled based on the presence of a vehicle that is detected with a low-power sensor in order to reduce the power consumption. Similarly, the camera-based systems may be activated only when the light condition is met in coordination with the light sensor, and different kinds of monitoring systems such as the infrared sensor based system can be activated at night. To the best of our knowledge, no research has been performed that identifies the optimal method for integrating various classification systems together. We envision that this review paper will be useful resources for development of such collaborative systems.

With rapid development of vehicle-to-everything (V2X) technology, we will see a mix of vehicles equipped with V2X devices and traditional ones on highways in the very near future. Traffic monitoring systems will have to provide support for classifying these V2X-equipped vehicles. Fortunately, classifying V2X-equipped vehicles will be very easy by allowing these vehicles to send vehicle class information via a V2X message to the classification system. Therefore, the classification accuracy will be significantly enhanced. However, numerous technical challenges for creating an effective protocol that enables seamless communication between passing cars and traffic monitoring systems will have to be addressed, such as reliable and secure data transmission, dynamic range adjustment, interference reduction, support for both DSRC and cellular network-based V2V, message formats, \emph{etc.}

\section{Conclusion}
\label{sec:conclusion}

We presented a review of traffic monitoring systems focusing on the key functionality of vehicle classification. By categorizing the vehicle classification systems according to how sensors are installed into three types, \emph{i.e.,} in-roadway, over-roadway, and side-roadway based systems, we discussed various research issues, methodologies, hardware design, and limitations. We also discussed a number of research challenges and future research directions. We expect that the rich contents about virtually all vehicle classification systems developed in the past decade will be useful resources for academia, industry, and government agencies in selecting appropriate vehicle classification solutions for their traffic monitoring applications.

\bibliographystyle{IEEEtran}
\bibliography{mybibfile}

\newpage

\begin{IEEEbiography}[{\includegraphics[width=1in,height=1.25in,clip,keepaspectratio]{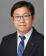}}]{Myounggyu Won} (M'13) received the Ph.D. degree in Computer Science from Texas A\&M University at College Station, in 2013. He is an Assistant Professor in the Department of Computer Science at the University of Memphis, Memphis, TN, United States. Prior to joining the University of Memphis, he was an Assistant Professor in the Department of Electrical Engineering and Computer Science at the South Dakota State University, Brookings, SD, United States from Aug. 2015 to Aug. 2018, and he was a postdoctoral researcher in the Department of Information and Communication Engineering at Daegu Gyeongbuk Institute of Science and Technology (DGIST), South Korea from July 2013 to July 2014.  His research interests include smart sensor systems, connected vehicles, mobile computing, wireless sensor networks, and intelligent transportation systems. He received the Graduate Research Excellence Award from the Department of Computer Science and Engineering at Texas A\&M University - College Station in 2012.
\end{IEEEbiography}

\EOD

\end{document}